\documentclass[12pt]{iopart}

\usepackage{iopams}

\expandafter\let\csname equation*\endcsname\relax

\expandafter\let\csname endequation*\endcsname\relax
\usepackage{amsmath}

\usepackage[mathlines]{lineno}
\usepackage{amssymb,latexsym,graphicx}
\usepackage{graphicx,units,multirow,array}
\newcolumntype{R}[1]{>{\raggedleft\let\newline\\\arraybackslash\hspace{0pt}}m{#1}}
\newcommand{\ab}{\nicefrac{\alpha}{\beta}}
\usepackage{color}
\usepackage{url}
\usepackage[abbr]{harvard}

\usepackage{soul}
\setstcolor{red}
\setul{}{2pt}

\usepackage{booktabs}

\begin{document}
\bibliographystyle{dcu}

\title[Spatiotemporal fractionation with bounds on the achievable benefit]{Optimization of spatiotemporally fractionated radiotherapy treatments with bounds on the achievable benefit}

\author{Melissa R. Gaddy}
\address{Department of Mathematics, North Carolina State University, Raleigh, NC 27695-8205, USA}
\author{Sercan Y\i ld\i z}
\address{Statistical and Applied Mathematical Sciences Institute, Research Triangle Park, NC 27709-4006, USA}
\author{Jan Unkelbach}
\address{Department of Radiation Oncology, University Hospital Z{\"u}rich, Z{\"u}rich, CH 8091 Switzerland}
\author{D\'{a}vid Papp}
\ead{dpapp@ncsu.edu} 
\address{Department of Mathematics, North Carolina State University, Raleigh, NC 27695-8205, USA}

\date{\today}

\begin{abstract}
Spatiotemporal fractionation schemes, that is, treatments delivering different dose distributions in different fractions, can potentially lower treatment side effects without compromising tumor control. This can be achieved by hypofractionating parts of the tumor while delivering approximately uniformly fractionated doses to the surrounding tissue. Plan optimization for such treatments is based on biologically effective dose (BED); however, this leads to computationally challenging nonconvex optimization problems. Optimization methods that are in current use yield only locally optimal solutions, and it has hitherto been unclear whether these plans are close to the global optimum. We present an optimization framework to compute rigorous bounds on the maximum achievable normal tissue BED reduction for spatiotemporal plans.

The approach is demonstrated on liver tumors, where the primary goal is to reduce mean liver BED without compromising any other treatment objective. The BED-based treatment plan optimization problems are formulated as quadratically constrained quadratic programming (QCQP) problems. First, a conventional, uniformly fractionated reference plan is computed using convex optimization. Then, a second, nonconvex, QCQP model is solved to local optimality to compute a spatiotemporally fractionated plan that minimizes mean liver BED, subject to the constraints that the plan is no worse than the reference plan with respect to all other planning goals. Finally, we derive a convex relaxation of the second model in the form of a semidefinite programming (SDP) problem, which provides a rigorous lower bound on the lowest achievable mean liver BED.

The method is presented on 5 cases with distinct geometries. The computed spatiotemporal plans achieve 12-35 percent mean liver BED reduction over the optimal uniformly fractionated plans. This reduction corresponds to 79-97 percent of the gap between the mean liver BED of the uniform reference plans and our lower bounds on the lowest achievable mean liver BED. The results indicate that spatiotemporal treatments can achieve substantial reductions in normal tissue dose and BED, and that local optimization techniques provide high-quality plans that are close to realizing the maximum potential normal tissue dose reduction.
\end{abstract}


\maketitle 

\section{Introduction}

Most radiotherapy treatments are fractionated, meaning that delivery of the total dose is split into multiple treatments delivered over several days or weeks. This is motivated by the \emph{fractionation effect}: the clinical observation that radiation-induced damage to cells is lower if the same physical dose is delivered over multiple fractions, owing to cells' ability to recover from sublethal radiation damage. The most widely used mathematical model of the fractionation effect is the biologically effective dose (BED) model \cite{Fowler2010}. According to this model, the biologically effective dose for a treatment with $N$ fractions, each delivering dose $d$, is given by
\begin{linenomath}
\begin{equation}
b = Nd\,\Big(1+\frac{d}{\ab}\Big), \label{eq:BED-basic}
\end{equation}
\end{linenomath}
where $\ab$ is a tissue-specific parameter. For a fixed total physical dose $Nd$, the BED is minimized if the dose is split evenly into many fractions. This standard fractionation regimen is desired in normal tissues. On the other hand, for a fixed total physical dose, the BED is maximized if all dose is delivered in few fractions (called \emph{hypofractionation}), which is desired in the tumor. Considering this inherent trade-off of fractionation decisions, it would seem ideal to simultaneously achieve hypofractionation in the tumor while splitting the dose to normal tissues evenly into many fractions. While this may appear unattainable at first glance, this goal can be achieved at least approximately using \emph{spatiotemporal fractionation schemes}.

\subsection{Spatiotemporal fractionation schemes}
Spatiotemporal fractionation schemes deliver different dose distributions in different fractions in an attempt to minimize BED in healthy tissue and maximize BED in the tumor by hypofractionating parts of the tumor while delivering approximately identical doses to the surrounding tissue. The clinical rationale for spatiotemporal fractionation to optimally exploit the fractionation effect was first proposed in the context of proton radiotherapy \cite{UnkelbachZengEngelsman2013,UnkelbachPapp2015}, where the potential benefit comes from the fact that the dose in the entrance region of a proton beam is largely independent of the beam's range, which provides some flexibility to modify the dose in the tumor without equally affecting the dose in the entrance region. More recently it has also been shown that spatiotemporal fractionation may provide a therapeutic advantage in arc therapy delivered with conventional photon beams \cite{Unkelbach2015}. In this case, arc therapy plans can be created in such a way that each fraction delivers high single-fraction doses to complementary parts of the target volume while creating a similar dose bath in the surrounding normal tissue. This was demonstrated for fractionated radiosurgery treatments of large arteriovenous malformations \cite{UnkelbachBussiereChapmanLoefflerShih2016}.

The optimization of spatiotemporally fractionated treatments is far more challenging than conventional IMRT/IMPT optimization. For most commonly used objective functions, the fluence map optimization problem in conventional IMRT planning is a large-scale convex optimization problem that can be solved to global optimality using well-established gradient-based (e.g., quasi-Newton) optimization methods \cite{Bortfeld2006}. Because spatiotemporal planning is based on BED rather than physical dose and the fluence maps for different fractions are distinct, the optimization models for optimal spatiotemporal fractionation are inherently nonconvex (see Section \ref{sec:nonconvexity}). It is common to use heuristics and local optimization methods for nonconvex models in RT planning, but the global optimality of the resulting plans is rarely discussed. In a recent work, \citeasnoun{AjdariGhate2016} studied a treatment planning problem of similar mathematical structure. They proposed a model predictive control approach which computes a treatment plan iteratively by optimizing the dose distributions of remaining fractions after each fraction, with the assumption that the same dose is delivered in every fraction thereafter. This approach requires only the solution of convex optimization problems; however, it does not promote the type of solutions that we seek in spatiotemporal fractionation: plans that deliver high doses to complementary parts of the tumor in distinct fractions. Furthermore, the model predictive control approach comes with no guarantees that the computed plans are anywhere close to globally optimal.

\subsection{The contribution of this paper}
So far there have been no computational tools to bound the maximum achievable benefit from spatiotemporal fractionation. Prior works \cite{UnkelbachPapp2015,UnkelbachBussiereChapmanLoefflerShih2016} use local optimization, and it has been unclear if better optimization algorithms may yield substantially better solutions.
Our approach combines the local optimization of the nonconvex treatment planning model for spatiotemporal fractionation with the solution of a convex optimization problem that provides a rigorous bound on the maximum achievable benefit from spatiotemporal fractionation. We approach this by formulating the spatiotemporal treatment planning problem as a nonconvex quadratically constrained quadratic programming (QCQP) problem. We then derive a convex relaxation of the QCQP problem in the form of a semidefinite programming (SDP) problem. The SDP relaxation provides a nontrivial lower bound; in particular, this bound is tighter than what can be achieved by replacing each nonconvex constraint in the QCQP model with its convex relaxation. We test our method on two-dimensional slices of 5 liver tumors that represent a variety of patient geometries. Comparing the quality of the locally optimal solutions against the SDP relaxation bounds, we find that the local optimal solutions computed for these cases are indeed nearly globally optimal. While there is no conceptual difficulty in applying the same approach to three-dimensional cases, and in particular the local optimal solutions can be computed using the same methods and software that we used in this paper, the SDP relaxations of the three-dimensional cases cannot be solved in a reasonable amount of time with available off-the-shelf software. 

\subsection{Relation to prior works}
Prior research has addressed the problem of optimizing fractionation decisions based on the BED model \cite{Mizutaetal2012,UnkelbachCraftSalariRamakrishnanBortfeld2013,KellerHopeMeierDavison2013,GayEtAl2013,SaberianGhateKim2016,SaberianGhateKim2015,MizutaDateEtAl2012}. These works aim at maximizing the tumor BED subject to BED constraints to the normal tissue. It was shown that the optimal number of fractions depends not only on the $\ab$ ratios of tumor and normal tissues, but also on the dose distribution. These works, however, all assume uniform fractionation, where the same dose distribution is delivered in all fractions, and only the number of fractions needs to be optimized.

The novelty in the idea of spatiotemporal fractionation lies in the fact that there is a potential advantage of delivering distinct dose distributions in different fractions, purely motivated by the basic fractionation effect as described by the standard BED model. There are several extensions of the BED model that describe higher order biological effects such as incomplete repair of radiation damage between fractions, repopulation of tumors over the course of treatment, accelerated repopulation effects, the effect of chemotherapeutic agents, and reoxygenation of hypoxic tumors \cite{HallGiaccia2012}. It was found that some of these models give rise to more complex fractionation schemes, i.e. varying doses per fraction \cite{BertuzziBruniPapaSinisgalli2013,BortfeldRamakrishnanTsitsklisUnkelbach2015,SalariUnkelbachBortfeld2015,WeinCohenWu2000,YangXing2005}. However, the role of such models to guide fractionation decisions in clinical practice has been limited. Instead, spatiotemporal fractionation as described in this paper is purely based on the basic fractionation effect, whose existence and clinical relevance is undoubted.

Another approach in which different dose distributions may arise in each fraction is \emph{adaptive radiotherapy} (ART) \cite{YanViciniWongMartinez1997,LuChenChenRuchalaOlivera2008,KimGhatePhillips2012}. This technique uses feedback information, such as changes in the patient anatomy, to modify the treatment plan during the course of a fractionated treatment, e.g. replanning to compensate for tumor shrinkage in lung or head and neck cancer \cite{SonkeBelderbos2010,WuChiChenKraussYanMartinez2009}. Instead, spatiotemporal fractionation shows that there is a benefit of delivering distinct dose distributions in different fractions, even in the absence of any changes of the patient over time.

\section{Mathematical model for optimal spatiotemporal fractionation}\label{sec:model}
\subsection{Uniform and nonuniform fractionation using the BED model}
The natural generalization of the BED model in \eqref{eq:BED-basic} to fractionated treatments delivering different doses in different fractions is to define the (cumulative) BED as
\begin{linenomath}
\begin{equation}
b = \sum_{t=1}^N \Big(d_t + \frac{d_t^2}{(\ab)}\Big), \label{eq:BED}
\end{equation}
\end{linenomath}
where $d_1,\dots,d_N$ are the doses delivered (in any order) in fractions $1$ through $N$.
IMRT planning using BED can be performed analogously to conventional IMRT optimization, using similar objective functions and constraints in the treatment planning optimization model, but substituting BED in place of physical dose \cite{UnkelbachPapp2015}. We define $x_1,\ldots,x_N$ to be the vectors of beamlet weights delivered in fractions $1$ through $N$, and $V$ to be the set of volume elements (voxels) used for dose calculation during the optimization. Using the cumulative BED from \eqref{eq:BED}, we obtain the nonuniform fractionation problem:
\begin{linenomath}
\begin{equation}\label{eq:IMRT-BED}
\begin{alignedat}{2}
\min_{x,d,b}\quad  & F(b)\\
\text{s.t.}\quad\; & b_v = \sum_{t=1}^N (d_{vt} + \tfrac{d_{vt}^2}{(\ab)_v} ) &\qquad& \forall\, v\in V \\
& Dx_t = d_t & & t=1,\dots, N\\
& x_t \geq 0 & & t=1,\dots, N,
\end{alignedat}
\end{equation}
\end{linenomath}
where $D$ is the usual dose-influence matrix,
and the objective function $F$ represents the desired clinical goals of target coverage, conformity, and organ sparing. Defining $I$ as the index set of the objectives, we write the objective function as the sum
\begin{linenomath}
\begin{equation}\label{eq:obj fun sum}
F(b) = \sum_{i\in I} w_iF_i(b),
\end{equation}
\end{linenomath}
where each term $F_i$ represents a single clinical objective (such as deviation from a prescribed lower or upper bound on the minimum, maximum, or mean BED of a structure), and each positive weight $w_i$ represents the relative importance of a clinical goal.

BED-based optimization can also be performed analogously for conventional, uniformly fractionated treatments. Throughout this work, we use a \emph{uniform reference plan} as a benchmark to evaluate the benefit of nonuniform fractionation. We obtain the uniform reference plan by solving \eqref{eq:IMRT-BED} with the additional constraint that $x_1=\cdots=x_N$. Eliminating the redundant variables, the uniform reference plan is the optimal solution of the following problem:
\begin{linenomath}
\begin{equation}\label{eq:IMRT-BED_uni}
\begin{split}
\min_{x,d,b}\quad  & F(b)\\
\text{s.t.}\quad\; & b_v = Nd_v\left(1+\frac{d_v}{(\ab)_v}\right) \qquad \forall\, v\in V\\
                   & Dx = d \\
                   & x\geq 0.
\end{split}
\end{equation}
\end{linenomath}

Figure \ref{fig: case1} shows an example of a reference plan and a nonuniform 5-fraction treatment of a patient harboring a large liver tumor. The reference plan (Figure 1c) was computed by solving the BED-based treatment plan optimization model \eqref{eq:IMRT-BED_uni}. Figure 1a shows the 5 dose distributions of a nonuniformly fractionated plan. Both plans were computed by optimizing the same objectives described below in Section \ref{sec:SDP-bound} to represent identical clinical goals. The reference plan yields a conventional treatment that irradiates the tumor to the prescribed dose in each fraction. The nonuniformly fractionated plan delivers high single-fraction doses to parts of the tumor while delivering a similar low-dose bath to the surrounding tissue in each fraction. The high single-fraction doses in the tumor allow for a reduction in the total physical dose delivered (Figure \ref{fig: case1}(b)). Since the dose to the surrounding normal tissue is approximately uniformly fractionated, this also yields a reduction of BED in the normal tissue.

\begin{figure*}[h]
\begin{tabular}{c c c c}
\multicolumn{4}{c}{(a)} \\
\multicolumn{4}{c}{\includegraphics[width=.8\textwidth]{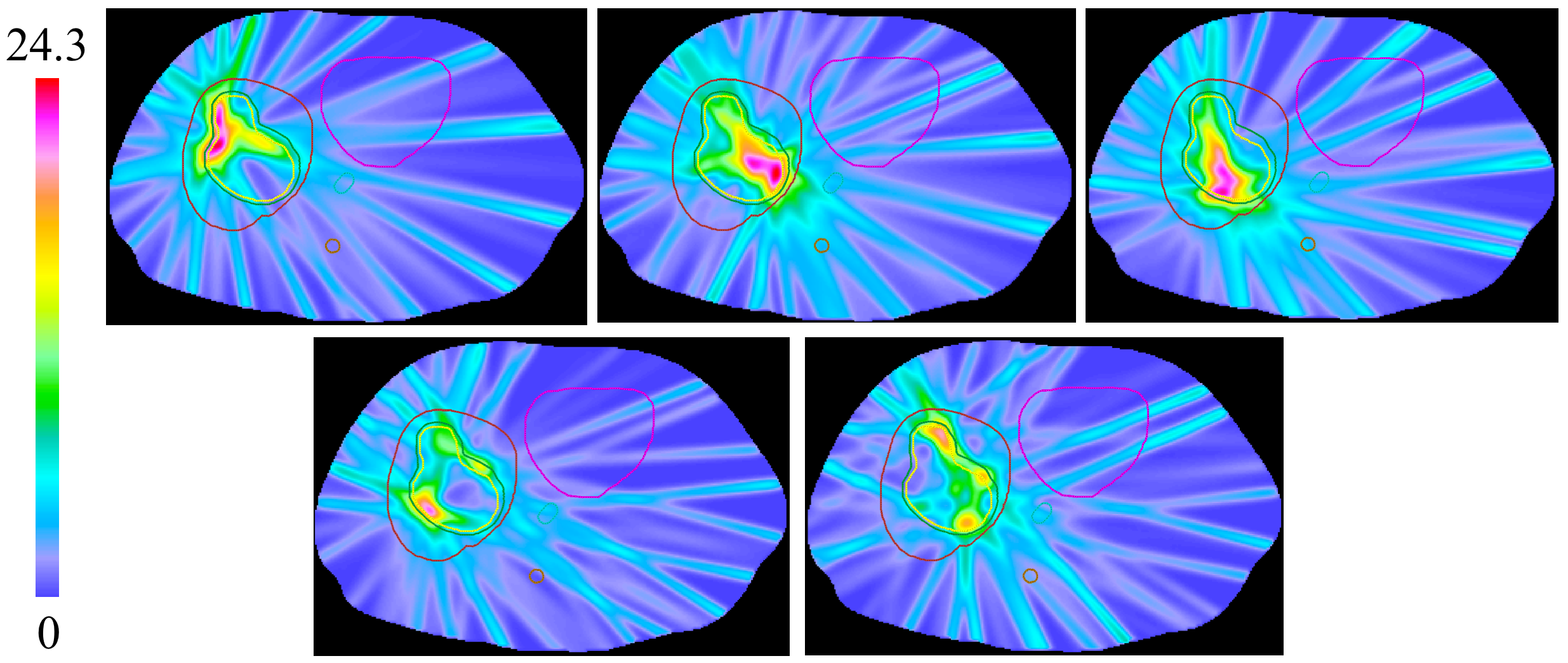}} \\
(b) & (c) & (d) & (e) \\
\includegraphics[width=.25\textwidth]{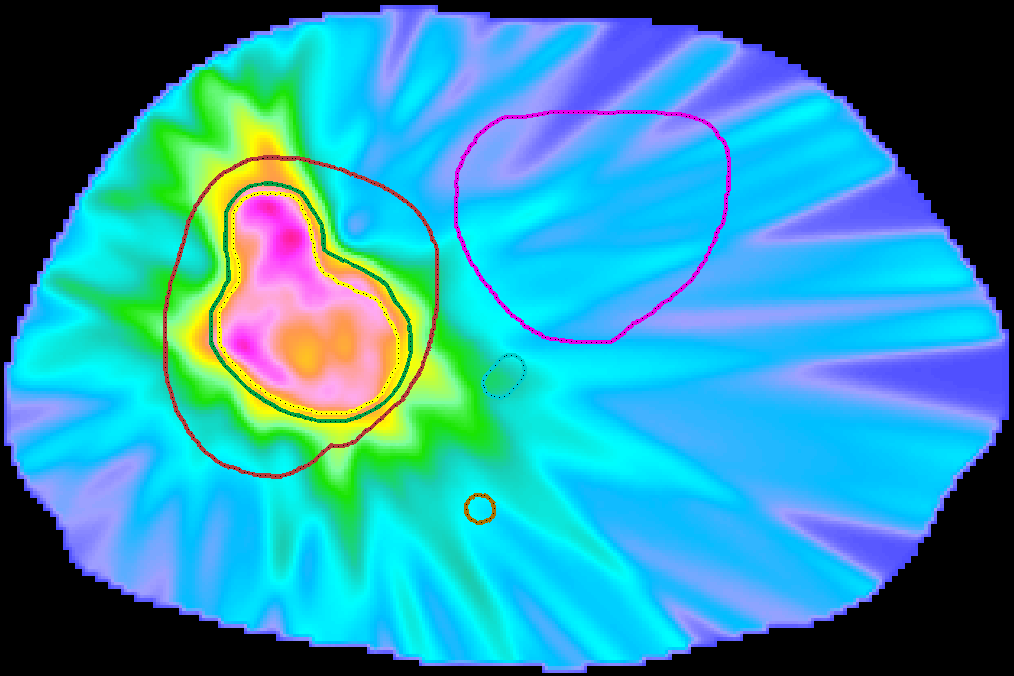} &
\includegraphics[width=.25\textwidth]{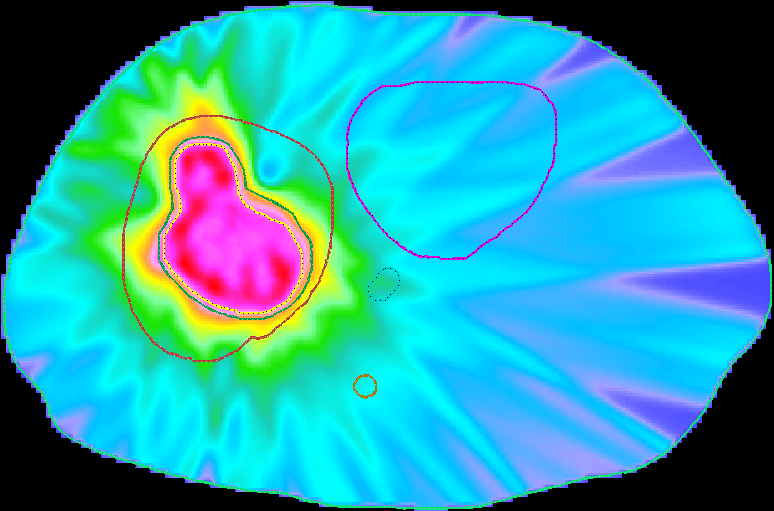} &
\includegraphics[width=.25\textwidth]{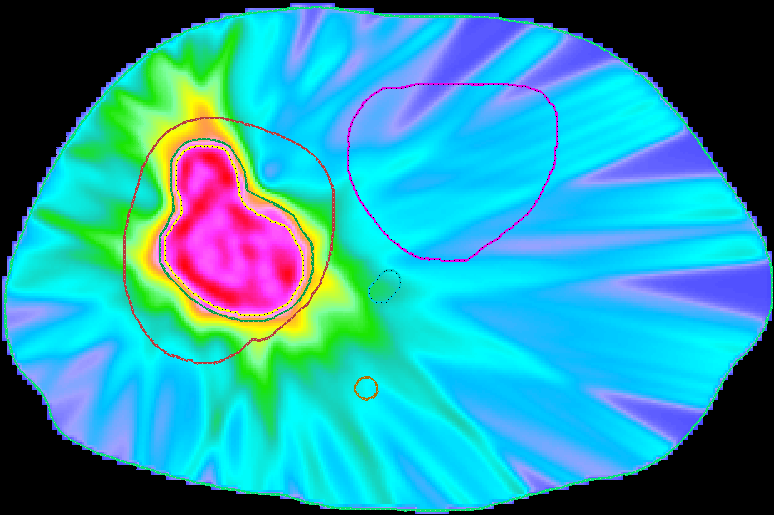} &
\includegraphics[width=.25\textwidth]{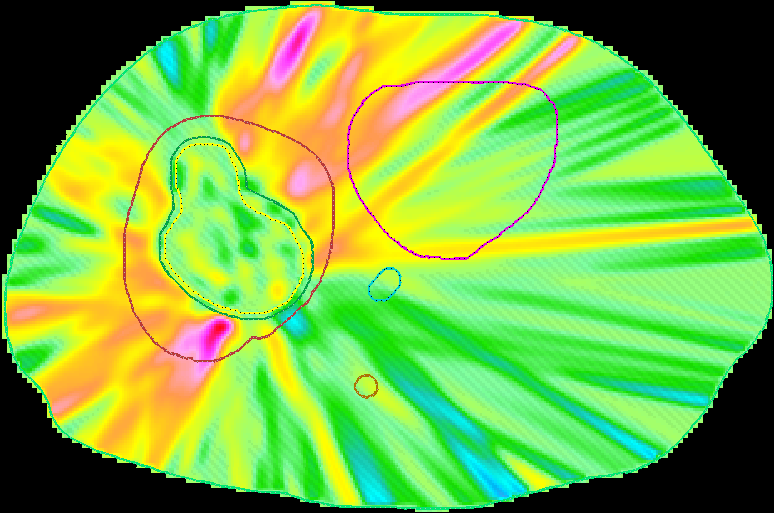} \\
\multicolumn{3}{c}{\includegraphics[height=.5cm]{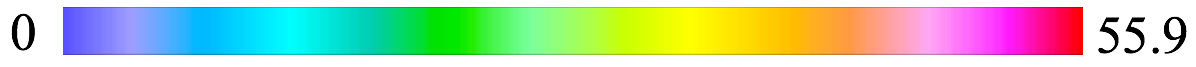}} &
\includegraphics[height=.45cm]{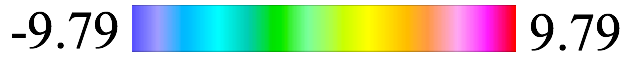} \\
\end{tabular}
\caption{Dose distributions for Case 1, a large central lesion within the liver. Also shown are the contours of the heart, the esophagus, and the spinal cord; these organs are not dose-limiting. (a) Physical dose distributions in each of the five fractions show that the nonuniformly fractionated treatment hypofractionates different parts of the tumor. (b) Total physical dose delivered throughout the nonuniformly fractionated treatment. (c) Physical dose distribution of the uniformly fractionated reference plan. (d) DEQ5 of the nonuniformly fractionated plan, which is the uniform plan that is isoeffective in delivering the same BED as the nonuniformly fractionated plan. (e) The difference between the physical dose in the uniform plan and the DEQ5 for the nonuniform plan, or (c) minus (d). This shows that the spatiotemporal plans reduce dose in the healthy liver and in the entrance region of the beams that expose the liver the most. All numerical quantities shown are in [Gy].}
\label{fig: case1}
\end{figure*}

\subsection{Nonconvexity of nonuniform fractionation}\label{sec:nonconvexity}

It is well-known that conventional fluence map optimization is a convex optimization problem
when the clinical goals are modeled using a convex objective function $F$ of the physical dose. Despite the apparent nonconvexity introduced by the quadratic equality constraints in the uniform fractionation problem \eqref{eq:IMRT-BED_uni}, this model is convex when the physical doses $d$ are restricted to clinically relevant values and $F$ is a piecewise quadratic penalty function similar to the dose-based objective functions \cite{UnkelbachPapp2015}. This is the case for the objectives used in this paper. Hence, the uniform reference plan can be computed using the gradient-based local optimization methods commonly used in IMRT plan optimization.

\begin{figure*}[h]
	\centering
\includegraphics[width=.9\textwidth]{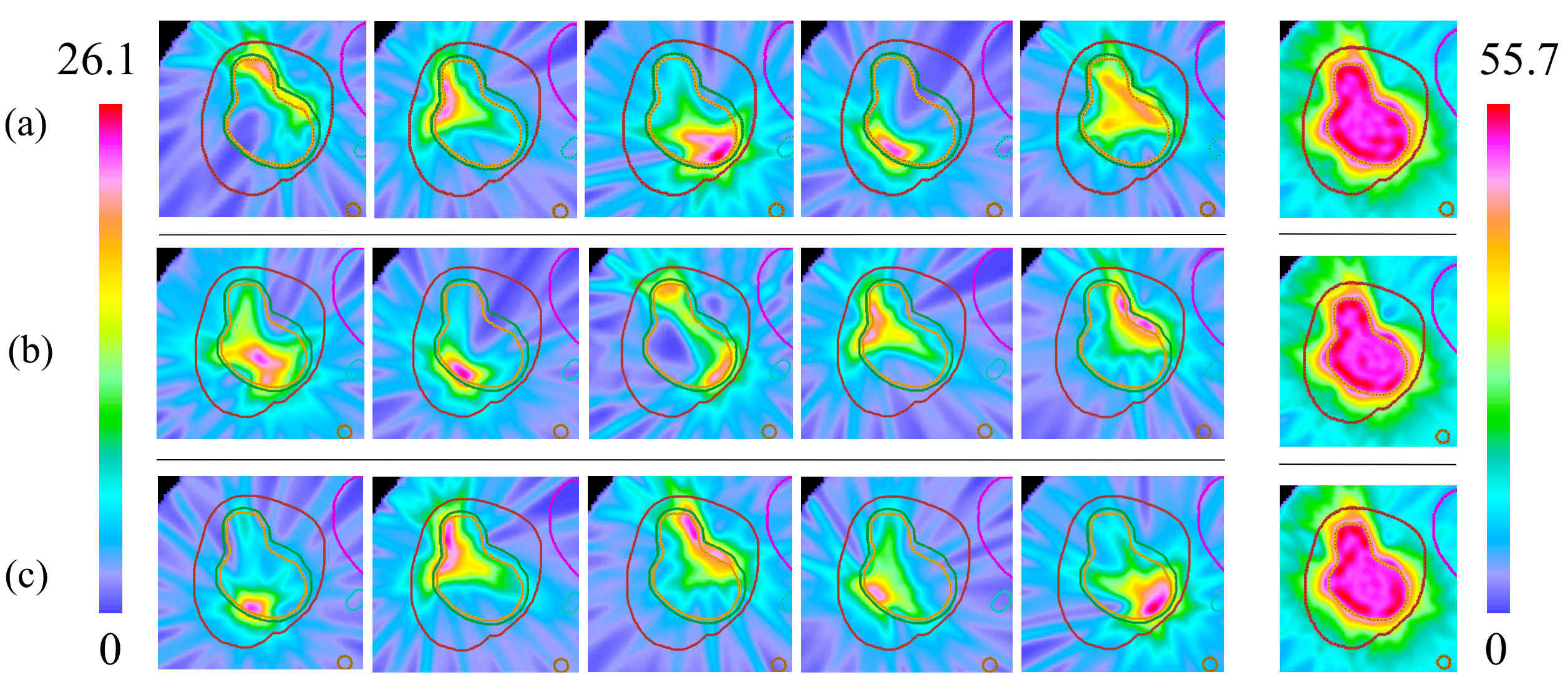}
\caption{Rows (a), (b), and (c) are physical dose distributions from three additional locally optimal 5-fraction nonuniform treatments for Case 1, which is the same clinical liver case displayed in Figure \ref{fig: case1}.  They are all locally optimal solutions of the model \eqref{eq:IMRT-BED}. The first five panels of each row are the dose distributions in the five nonuniform fractions, and the last panel on each row is the equivalent dose DEQ5 (see Section \ref{sec:results_comparison}). The solutions exhibit the same pattern: different subregions of the tumor receive a high single-fraction dose in different fractions. Note that the emergent ``partitions'' in the optimized plans are a result of the optimization. This pattern supports the rationale that the benefit of spatiotemporal fractionation is a result of hypofractionating parts of the tumor while maintaining a consistent low dose in the surrounding tissue. The difference in the hypofractionated regions in each solution also demonstrates that several qualitatively different locally optimal spatiotemporal treatments may exist for the same case.
All numerical quantities shown are in [Gy].}
\label{fig:local_solns}
\end{figure*}

The nonuniform fractionation problem \eqref{eq:IMRT-BED}, however, is nonconvex.\footnote{The nonconvexity of the formulation comes from the composition of the piecewise quadratic penalty function penalizing the underdose of the target and the function defining the BED. As a general rule, the composition of a convex nonincreasing function (such as our penalty function) with a convex quadratic function (such as the BED) can be nonconvex, unlike the composition of a convex nondecreasing function (such as the functions used to penalize overdose) and the BED, which is always convex.} This nonconvexity is an inherent characteristic of the problem that cannot be eliminated by reformulating the model, as the following argument shows: by definition, permuting the fractions in an optimal treatment plan leads to another optimal plan, yet the average of these $N!$ treatment plans will be a plan with identical fractions, which in general will not be optimal. Thus, gradient-based optimization methods used to solve the uniform model can only yield locally optimal solutions for the nonuniform model. \mbox{Figure \ref{fig:local_solns}} illustrates this problem, showing that substantially different locally optimal spatiotemporal plans may exist for the same case.

\subsection{The constrained nonuniform model}\label{sec:constr}
A drawback of formulation \eqref{eq:IMRT-BED}--\eqref{eq:obj fun sum} is that the optimal solution does not realize the maximum benefit of spatiotemporal fractionation over conventional fractionation in any given objective. Instead, the benefit is distributed among the terms of the objective function; in other words, the benefit is realized as a combination of smaller improvements with respect to the different clinical goals. In the liver cases we study in Section \ref{sec:results}, the primary gain is expected to be in lowering BED to the liver without compromising target coverage and conformity. This suggests an alternative formulation in which we minimize only one of the clinical objectives of \eqref{eq:obj fun sum}, say $F_1$, while constraining the solution to be at least as good as the uniform reference plan with respect to each other objective $F_i$ ($i\in I, i\neq 1$).

Let $b^*$ be the BED distribution delivered by the uniform reference plan, which is obtained by solving \eqref{eq:IMRT-BED_uni}, and let $F_1$ be the penalty for the mean BED in the liver exceeding zero. We modify \eqref{eq:IMRT-BED} to include the constraints that the solution must be as good as the uniform reference plan with respect to all objectives besides $F_1$, and we obtain the \emph{constrained} nonuniform spatiotemporal fractionation problem:
\begin{linenomath}
\begin{equation}\label{eq:IMRT-BED_constr}
\begin{alignedat}{2}
\min_{x,d,b}\quad  & F_1(b)\\
\text{s.t.}\quad\; & F_i(b) \leq F_i(b^*)  & &i\in I,\; i\neq 1 \\
			       & b_v = \sum_{t=1}^N (d_{vt} + \tfrac{d_{vt}^2}{(\ab)_v} ) & \qquad & \forall\,v\in V\\
                   & Dx_t = d_t & & t=1,\dots, N\\
                   & x_t \geq 0 & & t=1,\dots, N.
\end{alignedat}
\end{equation}
\end{linenomath}
The first set of constraints ensures that the improvement in the objective $F_1$ is not at the cost of sacrificing the other clinical objectives; the computed spatiotemporal plan is either preferable or identical to the uniform reference plan with respect to every objective.

\section{Mathematical model for bounding the maximum achievable benefit}\label{sec:SDP-bound}

In this section, we formulate the convex optimization model for bounding the maximum achievable benefit from spatiotemporal fractionation. To that end, we write problem \eqref{eq:IMRT-BED}--\eqref{eq:obj fun sum} as a quadratic optimization problem with quadratic constraints. We assume that $F$ is a weighted sum \eqref{eq:obj fun sum} of penalty functions that penalize deviations from prescribed BED values. Let $V_i$ be the set of voxels involved in clinical objective $i\in I$. Each $F_i$ is a piecewise quadratic penalty function that penalizes either BED above a prescribed threshold $b_v^{hi}$ in voxels $v\in V_i$, or BED below a prescribed threshold $b_v^{lo}$ in voxels $v\in V_i$, or mean BED above a prescribed mean BED $m_i^{hi}$ in a structure $i$. Let $I^+$, $I^-$, and $I^m$ be the index sets of objectives of the first, second, and third type respectively. (Thus, the index set $I$ in \eqref{eq:obj fun sum} is the union of the three sets $I^+, I^-$, and $I^m$.) Letting $(y)_+$ denote the positive-part function $\max(y,0)$, we express the clinical goal $i\in I$ with the penalty function
\begin{linenomath}
\begin{equation}\label{eq:obj_Fi} F_i(b) =
\begin{cases}
\displaystyle\sum_{v\in V_i} (b_v - b_{iv}^{hi})_{+}^2&\text{ if }i\in I^+,\\
\displaystyle\sum_{v\in V_i} (b_{iv}^{lo} - b_v)_{+}^2&\text{ if }i\in I^-,\\
\displaystyle\bigg(\frac{1}{|V_i|}\sum_{v\in V_i} b_v - m_i^{hi}\bigg)_{+}^2&\text{ if }i\in I^m.
\end{cases}
\end{equation}
\end{linenomath}
The overall objective function $F$ is the weighted sum \eqref{eq:obj fun sum}.

The prescription BED values $b_{iv}^{hi}$ and $b_{iv}^{lo}$ can be derived from clinical dose prescriptions of fractionated treatments that employ similar physical doses per fraction as expected to be used in the spatiotemporally fractionated treatments. Typically one would use the same threshold $b^{hi}$ or $b^{lo}$ in every voxel $v$ of the same structure $V_i$, but we opted for the above, more general, formulation to allow for distance-dependent thresholds often used to improve the conformity of the dose distributions.

Noting that $(y)_{+}$ is the smallest number $z$ satisfying the inequalities $z \geq y$ and $z\geq 0$, we introduce the auxiliary optimization variables $p_{iv}$ for $i \in I^+$ and $q_{iv}$ for $i \in I^-$, whose values (when minimized) are equal to the quantities $(b_v - b_{iv}^{hi})_{+}$ and $(b_{iv}^{lo} - b_v)_{+}$ for voxel $v$. Similarly, we introduce the variables $r_i$ for $i \in I^m$ for the mean-BED penalties $(\frac{1}{|V_i|}\sum_{v\in V_i} b_v - m_i^{hi})_{+}$.

Finally, we eliminate the variables $b_v$ from the problem by replacing them with $\sum_{t=1}^N (d_{vt} +  \tfrac{d_{vt}^2}{(\ab)_v})$. We arrive at the following quadratically constrained quadratic programming (QCQP) formulation of the nonuniform fractionation problem:
\begin{linenomath}
\begin{equation}\label{eq:quadmodel}
	\begin{alignedat}{2}
	\min_{x,d,p,q,r}\quad & \sum_{i\in I^+} \sum_{v\in V_i} w_ip_{iv}^2 + \sum_{i\in I^-} \sum_{v\in V_i} w_iq_{iv}^2 + \sum_{i\in I^m} w_ir_i^2 \\
	\text{s.t.}\quad\;\;\, & p_{iv} \geq - b_{iv}^{hi} + \sum_{t=1}^N (d_{vt} + \tfrac{ d_{vt}^2}{(\ab)_v}) & & \forall i \in I^+,\; \forall v\in V_i\\
	& p_{iv} \geq 0 & & \forall i \in I^+, \forall v\in V_i \\
	& q_{iv} \geq b_{iv}^{lo} - \sum_{t=1}^N (d_{vt}+ \tfrac{ d_{vt}^2}{(\ab)_v}) & & \forall i \in I^-,\;\forall v\in V_i\\
	& q_{iv} \geq 0 & & \forall i \in I^-,\;\forall v\in V_i \\
	& r_i \geq \frac{1}{|V_i|} \sum_{v\in V_i} \left( \sum_{t=1}^N d_{vt} + \tfrac{ d_{vt}^2}{(\ab)_v} \right) - m_i^{hi} & \quad & \forall i \in I^m \\
	& r_i \geq 0 & & \forall i \in I^m \\
	& Dx_t = d_t & & \forall t = 1,\ldots, N \\
	& x_t \geq 0 & & \forall t = 1,\ldots, N.
	\end{alignedat}
\end{equation}
\end{linenomath}

As the physical dose $d_{vt}$ is a linear function of the beamlet weights $x_t$, the $d_{vt}$ variables can also be eliminated from the formulation, and all the inequality constraints can be seen as quadratic inequalities in the primary decision variables $x_1,\dots,x_N$. We introduce the matrix $X_t = x_t x_t^T$ and consider each matrix element as an auxiliary decision variable. The quadratic inequalities in \eqref{eq:quadmodel} can then be written as linear inequalities in $x_t$ and $X_t$. For instance, on the right-hand side of the first set of inequalities we substitute
\begin{linenomath}
\begin{equation*} -b_{iv}^{hi} + \sum_{t=1}^N (d_{vt} + \tfrac{ d_{vt}^2}{(\ab)_v})  = \sum_{t=1}^N \left\langle \begin{pmatrix} 1 & x_t^T \\ x_t & X_t \end{pmatrix} , \begin{pmatrix}  -\frac{b_{iv}^{hi}}{N} & \frac{e_v^TD}{2} \\ \frac{D^Te_v}{2} & \tfrac{ 1}{(\ab)_v} D^Te_ve_v^TD\end{pmatrix}\right\rangle,
\end{equation*}
\end{linenomath}
where $\langle A,B\rangle = \sum_{i,j}A_{ij}B_{ij}$ is the component-wise (or Frobenius) inner product of matrices, and $e_v$ is the characteristic vector with a 1 in the $v$-th position and zeros elsewhere. (That is, $e_v^TD$ is simply the $v$-th row of the $D$ matrix.) The resulting optimization model has a convex quadratic objective function and only linear constraints, aside from the nonconvex quadratic equations $X_t = x_t x_t^T$ for $t=1,\dots,N$. We obtain a convex relaxation of this problem by replacing each of these equations with the weaker convex constraint $X_t - x_t x_t^T \succcurlyeq 0$ (meaning that the difference $X_t - x_t x_t^T$ is positive semidefinite). The latter constraint is indeed convex, as it is equivalent to the linear matrix inequality $\left(\begin{smallmatrix} 1 & x_t^T \\ x_t & X_t \end{smallmatrix}\right) \succcurlyeq 0$.

Since in the original nonconvex model we have $X_t=x_tx_t^T$ and $x_t\geq 0$, it is clear that each $X_t$ is also component-wise nonnegative. Hence, we can add the inequalities $X_t\geq 0$ for $t=1,\ldots,N$ to the convex relaxation. (This component-wise inequality should not be confused with the linear matrix inequality $X_t\succcurlyeq 0$.) Note that even though the inequality $X_t\geq 0$ is redundant in the original model, $X_t\geq 0$ is not a redundant constraint in the convex relaxation; adding it to the optimization problem tightens the bound.

A further simplification is possible. Since the relaxation is convex and symmetric in the fractions, we can assume without loss of generality that at the optimum we have $x_1=\cdots=x_N$ and $X_1=\cdots=X_N$. Thus, we can eliminate the variables corresponding to the different fractions and use only a single variable $x$ and $X$ in place of each $x_t$ and $X_t$. (This shows that our convex relaxation does not distinguish between the uniformly and nonuniformly fractionated models, although the bound does depend on the number of fractions.)

Finally, using the shorthand $C_v = \tfrac{ 1}{(\ab)_v}D^Te_ve_v^TD$, we arrive at the following convex relaxation of \eqref{eq:quadmodel}:
\begin{linenomath}
\begin{equation}\label{eq:SDPrelaxation}
\begin{alignedat}{2}
\min_{x,X,p,q,r} \quad & \sum_{i\in I^+} \sum_{v\in V_i} w_ip_{iv}^2 + \sum_{i\in I^-} \sum_{v\in V_i} w_iq_{iv}^2 + \sum_{i\in I^m} w_ir_i^2 \\
\text{s.t.} \quad\;\;\; & p_{iv} \geq N \left\langle \begin{pmatrix} 1 & x^T \\ x & X \end{pmatrix} , \begin{pmatrix}  -\frac{b_{iv}^{hi}}{N} & \frac{e_v^TD}{2} \\ \frac{D^Te_v}{2} & C_v\end{pmatrix}\right\rangle & & \forall i \in I^+,\; \forall v\in V_i \\
&p_{iv} \geq 0 & & \forall i \in I^+,\; \forall v\in V_i \\
&q_{iv} \geq -N \left\langle \begin{pmatrix} 1 & x^T \\ x & X \end{pmatrix} , \begin{pmatrix}  -\frac{b_{iv}^{lo}}{N} & \frac{e_v^TD}{2} \\ \frac{D^Te_v}{2} & C_v\end{pmatrix}\right\rangle & & \forall i \in I^-,\; \forall v\in V_i \\
&q_{iv} \geq 0 & & \forall i \in I^-,\; \forall v\in V_i \\
& r_i \geq \frac{N}{|V_i|} \sum_{v\in V_i} \left\langle \begin{pmatrix} 1 & x^T \\ x & X \end{pmatrix} , \begin{pmatrix}  -\frac{m_i^{hi}}{N} & \frac{e_v^TD}{2} \\ \frac{D^Te_v}{2} & C_v\end{pmatrix}\right\rangle &\qquad & \forall i\in I^m \\
& r_i \geq 0 & & \forall i\in I^m \\
& \begin{pmatrix} 1 & x^T \\ x & X \end{pmatrix} \succcurlyeq 0,\;\; x \geq 0,\;\; X \geq 0.
\end{alignedat}
\end{equation}
\end{linenomath}
The optimal objective function value of this problem is a lower bound for the global minimum of the spatiotemporal fractionation problem \eqref{eq:quadmodel}.

Note that to obtain \eqref{eq:SDPrelaxation}, we linearized every quadratic constraint in \eqref{eq:quadmodel}, even though all of them were convex except for those involving $q_{iv}$. It can be seen that this way \eqref{eq:SDPrelaxation} yields a tighter bound than what could be obtained by simply replacing the concave quadratic constraints with a convex relaxation and keeping the convex quadratic constraints intact. This is because unlike the linearizations of the concave quadratics (which are relaxations), the linearizations of the convex quadratics are tighter than the original constraints. Our derivation of \eqref{eq:SDPrelaxation} shows that the convex model as a whole is indeed a relaxation of \eqref{eq:quadmodel} despite the fact that this cannot be seen when we compare the models constraint by constraint.

\subsection{Bounding the maximum benefit in a given objective}\label{sec:nonuniform-constrained}

The convex relaxation of the constrained nonuniform model \eqref{eq:IMRT-BED_constr} can be derived analogously to how we obtained the relaxation \eqref{eq:SDPrelaxation} of the model \eqref{eq:IMRT-BED}.
For simplicity of notation, we assume (consistently with the application to liver tumors) that the mean dose objective $F_1(b)=(\frac{1}{|V_1|}\sum_{v\in V_1} b_v - m_1^{hi})_{+}^2$ is the primary objective.
Then the convex optimization model bounding the minimum value of $F_1(b)$ from below is the following:
\begin{linenomath}
\begin{equation}\label{eq:SDPrelaxation-constrained}
\begin{alignedat}{2}
\min_{x,X,p,q,r} \quad & r_1 \\
\text{s.t.} \quad\;\;\; & \sum_{v\in V_i}  p_{iv}^2 \leq F_i(b^*) & & \forall i \in I^+\\
& \sum_{v\in V_i}  q_{iv}^2 \leq F_i(b^*) & & \forall i \in I^-\\
& r_i \leq F_i(b^*) & & \forall i \in I^m,\; i \neq 1\\
& p_{iv} \geq N \left\langle \begin{pmatrix} 1 & x^T \\ x & X \end{pmatrix} , \begin{pmatrix}  -\frac{b_{iv}^{hi}}{N} & \frac{e_v^TD}{2} \\ \frac{D^Te_v}{2} & C_v\end{pmatrix}\right\rangle & &\forall i \in I^+,\; \forall v\in V_i\\
&p_{iv} \geq 0 & &\forall i \in I^+,\; \forall v\in V_i\\
&q_{iv} \geq -N \left\langle \begin{pmatrix} 1 & x^T \\ x & X \end{pmatrix} , \begin{pmatrix}  -\frac{b_{iv}^{lo}}{N} & \frac{e_v^TD}{2} \\ \frac{D^Te_v}{2} & C_v\end{pmatrix}\right\rangle & &\forall i \in I^-,\; \forall v\in V_i\\
&q_{iv} \geq 0 & &\forall i \in I^-,\; \forall v\in V_i \\
& r_i \geq \frac{N}{|V_i|} \sum_{v\in V_i} \left\langle \begin{pmatrix} 1 & x^T \\ x & X \end{pmatrix} , \begin{pmatrix}  -\frac{m_i^{hi}}{N} & \frac{e_v^TD}{2} \\ \frac{D^Te_v}{2} & C_v\end{pmatrix}\right\rangle &\quad & \forall i\in I^m \\
& r_i \geq 0 & & \forall i\in I^m \\
& \begin{pmatrix} 1 & x^T \\ x & X \end{pmatrix} \succcurlyeq 0,\;\; x \geq 0,\;\; X \geq 0.
\end{alignedat}
\end{equation}
\end{linenomath}

\subsection{Solution methods}

Although the optimization problems \eqref{eq:SDPrelaxation} and \eqref{eq:SDPrelaxation-constrained} are convex, their solution is not straightforward using the gradient-based algorithms commonly used in treatment planning optimization, such as the projected gradient descent method and its variants. This is because computing the violation of the linear matrix inequality constraints $\left(\begin{smallmatrix} 1 & x^T \\ x & X \end{smallmatrix}\right) \succcurlyeq 0$ and projecting on the set of points satisfying this inequality require an expensive matrix factorization in each step of the method. Optimization problems involving this type of constraints are called \emph{semidefinite programs} \cite{VandenbergheBoyd1996,BoydVandenberghe2004}. They have been extensively studied in the numerical optimization literature, and several reliable and efficient algorithms are available for their solution.

Formally, a semidefinite program is a convex optimization problem with a linear objective function and with constraints that are all either linear inequalities or \emph{linear matrix inequalities} of the form $A(x)\succcurlyeq 0$, where $A(\cdot)$ is an affine function that maps the optimization variables $x$ to the space of real symmetric matrices. The problems \eqref{eq:SDPrelaxation} and \eqref{eq:SDPrelaxation-constrained} have convex quadratic objectives and constraints, rather than only linear ones, but it can be shown that convex quadratic inequalities can be equivalently written as linear matrix inequalities \cite{VandenbergheBoyd1996}; hence their inclusion is without loss of generality.

Large-scale semidefinite programs are routinely solved using Newton-type methods implemented in widely available optimization software such as SeDuMi \cite{sedumi} and MOSEK \cite{mosek}. In our experiments we used MOSEK to solve the semidefinite programs \eqref{eq:SDPrelaxation} and \eqref{eq:SDPrelaxation-constrained}.

\section{Application to liver tumors}\label{sec:results}
We examined the benefit of spatiotemporal fractionation on two-dimensional slices of five clinical liver cases with distinct geometries. The first three cases feature centrally-located lesions within the liver; Case 1 has a large lesion, Case 2 has a small lesion, and Case 3 has two separate lesions within the liver. In each of these cases the liver is the main dose-limiting organ. In Case 4, the tumor abuts the chest wall, and Case 5 is a challenging geometry where both the chest wall and the bowel are dose-limiting and need to be included in the treatment plan optimization model.

\subsection{Experimental setup}

\subsubsection{The BED prescription and $\ab$ ratios.}
Five-fraction treatments were planned for all cases. To derive the upper and lower BED thresholds $b^{hi}$ and $b^{lo}$, typical prescription doses and normal tissue constraints for 5-fraction liver SBRT were converted into BED values using an $\ab$ ratio of 10 in the tumor and 4 in all normal tissues. For example, the prescription lower bounds $b^{lo}$ for the GTV and PTV were chosen to be 100 Gy and 72 Gy, respectively, which correspond to 50 Gy and 40 Gy of physical dose delivered in 5 fractions assuming uniform fractionation. The complete list of objectives and constrains is as follows:
\begin{itemize}
	\item A BED of 100 Gy is prescribed to the GTV. 
	\item A BED of 72 Gy is prescribed to the PTV.
	\item BED exceeding 115.5 Gy BED in the GTV is penalized.
	\item BED exceeding 100 Gy BED in the PTV is penalized.
	\item The plan is to be conformal: a linear BED falloff is aimed for in a 3cm margin around the PTV.
	\item The mean BED in the liver excluding the GTV is minimized.
	\item In Cases 4 and 5, BED exceeding 96.2 Gy in the chest wall is penalized. (This corresponds to 35 Gy physical dose in 5 fractions.)
	\item In Case 5, BED exceeding 75 Gy in the affected sections of the GI tract is penalized. 
\end{itemize}	
Each of these objectives is implemented via a penalty function shown in Equation \eqref{eq:obj_Fi}.

\subsubsection{Computing the uniformly fractionated reference plan.}
\label{sec:results-uniform}

The BED-based fluence map optimization problem \eqref{eq:IMRT-BED_uni} was solved to obtain optimal beamlet weights for the 5-fraction uniform reference plan, which is a high-quality plan that delivers the same dose distribution in each of its fractions. Dose-influence matrices were calculated for 21 equispaced coplanar beams using the Quadrant Infinite Beam (QIB) dose calculation algorithm implemented in CERR version 5.2 \cite{cerr}. These 21-beam IMRT plans represent the quality achievable by high-quality arc therapy plans \cite{PappUnkelbach2014}. In this work, we limited the computations to two-dimensional slices of the patient voxels and optimized beamlet weights for a single row of the 1x1 cm beamlet grid. As discussed in Section \ref{sec:nonconvexity}, although the uniform plan optimization problem is defined using a combination of convex and nonconvex constraints, when the beamlet weights $x$ and physical doses $d$ are restricted to clinically relevant values, the problem is convex, and the global optimal solution can be computed using local optimization algorithms \cite{UnkelbachPapp2015}. In our experiments, the optimization of the reference plans was performed with Matlab using the L-BFGS-B solver \cite{lbfgsb-web}. The runtimes to find the (globally) optimal solutions were less than 10 seconds on a standard desktop computer for all of the cases.

\subsubsection{Computing the spatiotemporally fractionated plans.}
After computing the uniform reference plan, we computed nonuniformly fractionated treatment plans in which the beamlet weights and corresponding dose distributions are not the same in each of the five fractions. As described in Section \ref{sec:constr}, we solve \eqref{eq:IMRT-BED_constr} to minimize the mean BED in the liver, subject to the constraints that the solution must be at least as good as the uniform reference plan computed in Section \ref{sec:results-uniform} with respect to all other objectives. A perturbation of the fluence maps of the uniform reference plan was used as an initial solution for the optimization.\footnote{Beamlet weights were multiplied by independent random factors drawn uniformly from the interval $[0,2]$. The quality of the computed local optimal solutions did not appear to be sensitive to the specific way the plans were perturbed.} These nonconvex problems were solved to local optimality using Matlab's built-in optimizer \texttt{fmincon} \cite{fmincon}, which utilizes an interior-point algorithm to find local solutions to a constrained nonlinear program. Runtimes for computing a locally optimal solution of \eqref{eq:IMRT-BED_constr} ranged from 15 minutes to 4 hours.

\subsection{Results}\label{sec:results_comparison}
\begin{figure*}[h]
\begin{tabular}{c c c c}
\multicolumn{4}{c}{(a)} \\
\multicolumn{4}{c}{\includegraphics[width=.8\textwidth]{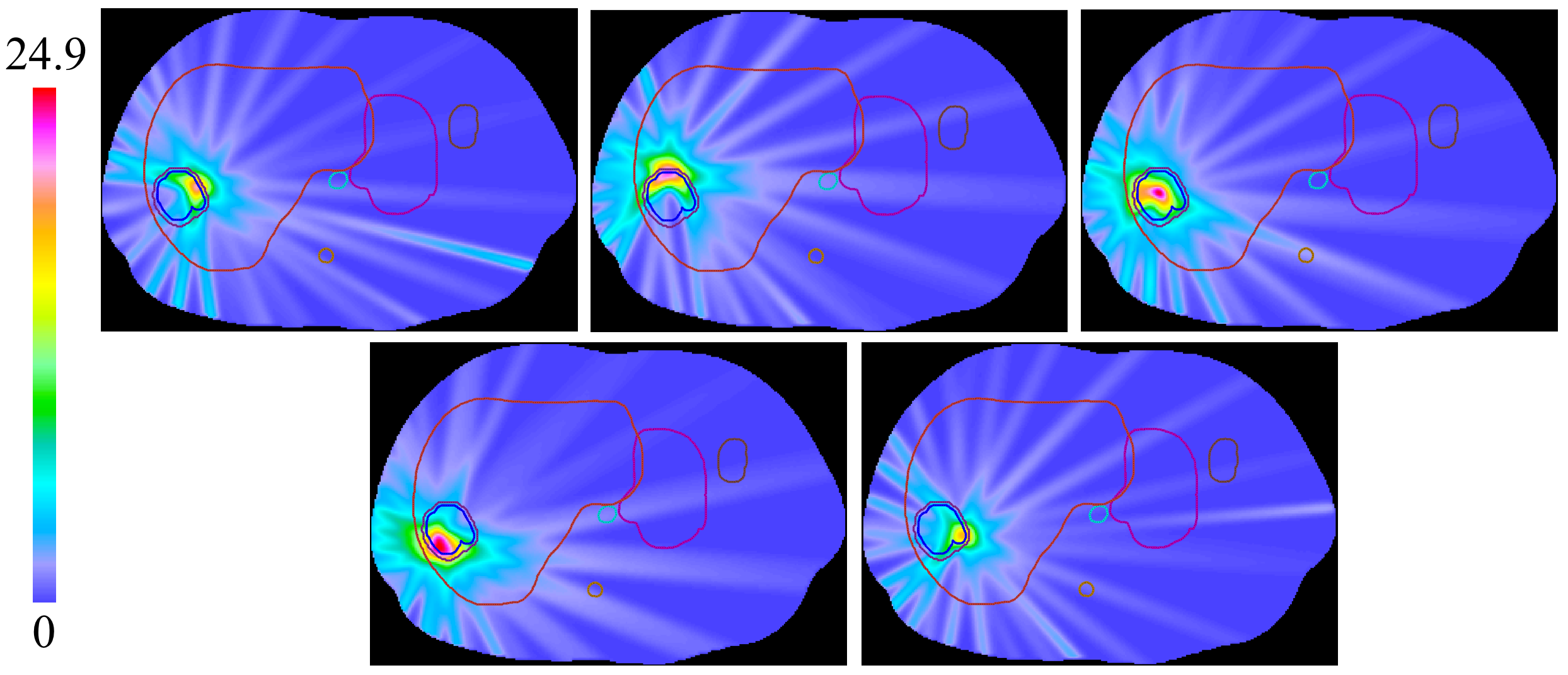}} \\
(b) & (c) & (d) & (e) \\
\includegraphics[width=.25\textwidth]{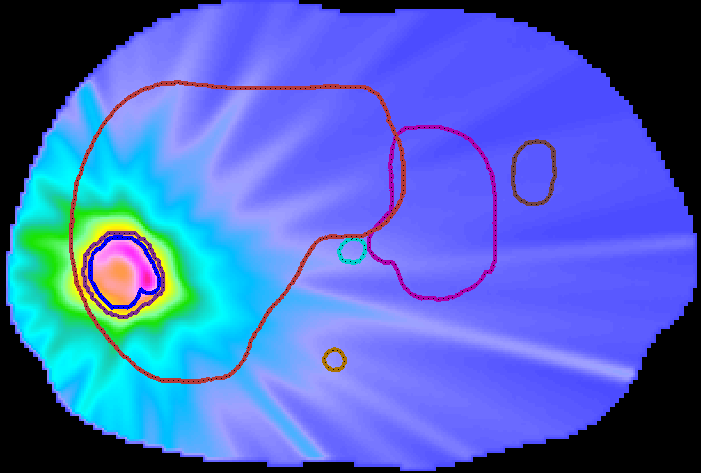} &
\includegraphics[width=.25\textwidth]{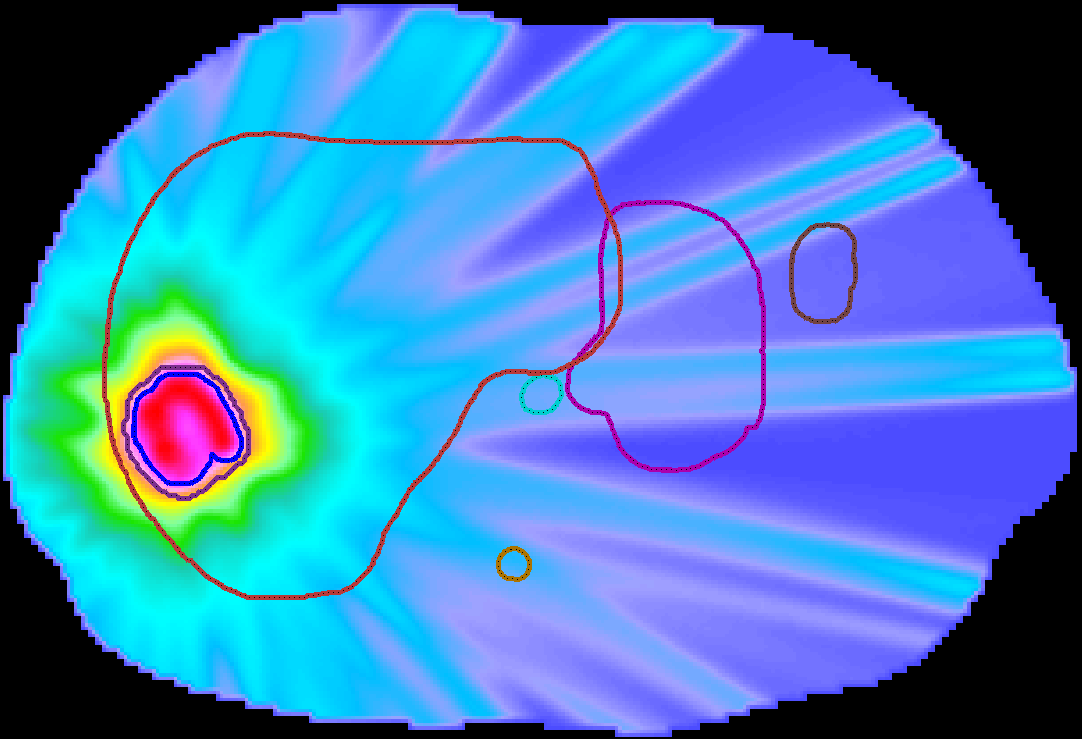} &
\includegraphics[width=.25\textwidth]{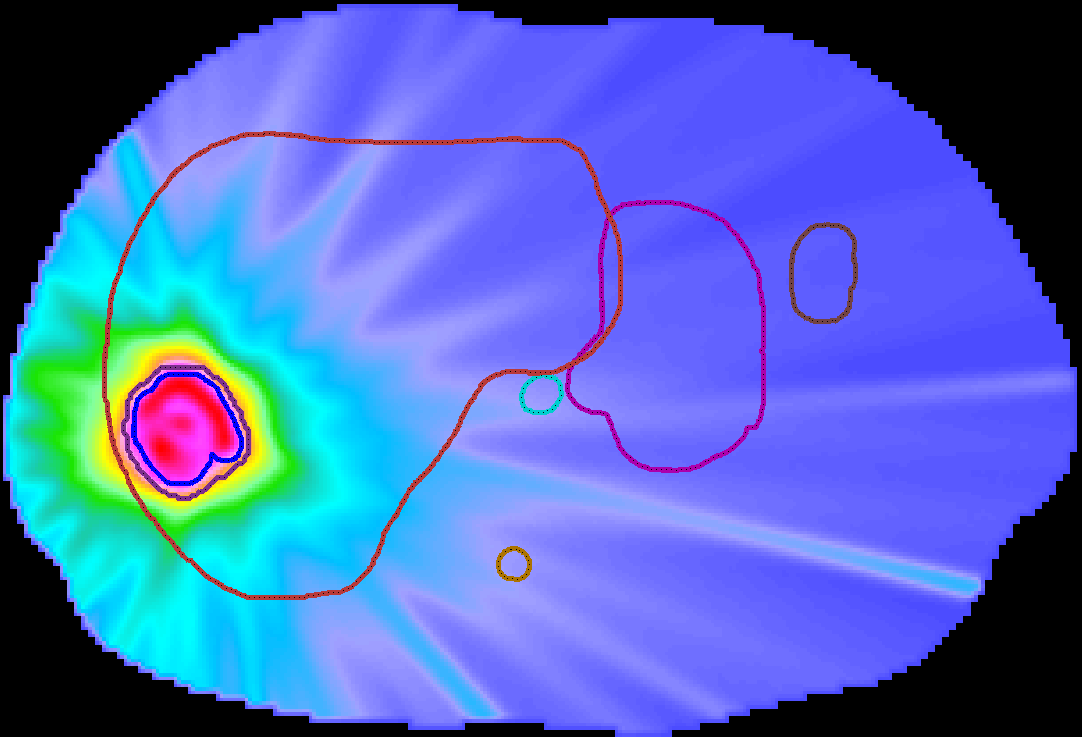} &
\includegraphics[width=.25\textwidth]{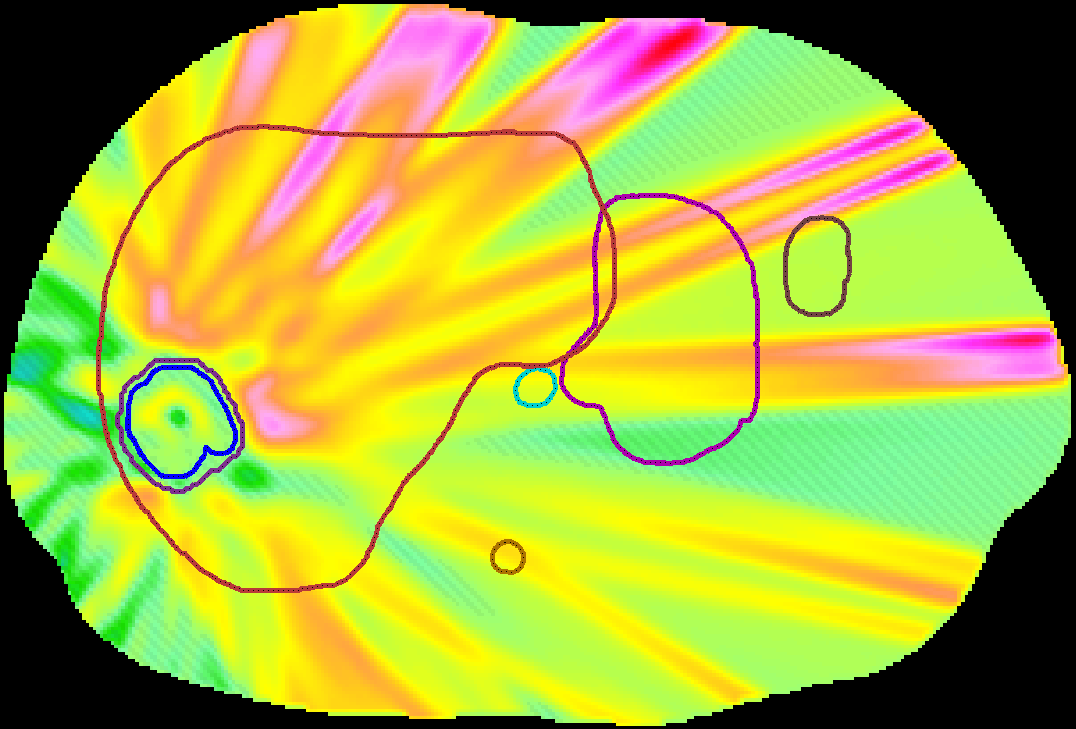} \\
\multicolumn{3}{c}{\includegraphics[height=.5cm]{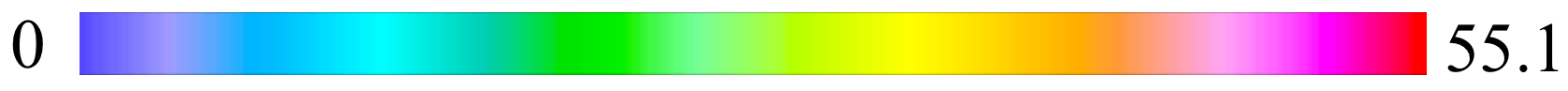}} &
\includegraphics[height=.45cm]{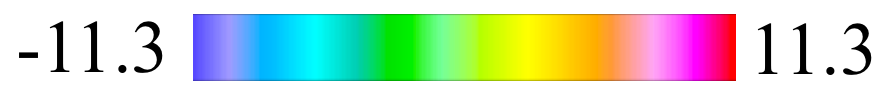} \\
\end{tabular}
\caption{Dose distributions for Case 2, a small lesion within the liver. (a) Physical dose distributions in a 5-fraction nonuniformly fractionated treatment. Although the tumor is small, the spatiotemporal plan still hypofractionates different subregions of the target. (b) Total physical dose delivered by the five nonuniform fractions. (c) Physical dose distribution of the uniform reference plan. (d) DEQ5 of the nonuniformly fractionated plan. As in Case 1, the nonuniformly fractionated plan provides similar target coverage as the uniform reference plan, with a more conformal dose distribution. (e) The plot of (c) minus (d) shows that the majority of the BED reduction is in the healthy liver and in the entrance regions of the beams exposing the liver the most. All numerical quantities shown are in [Gy].}
\label{fig: case2}
\end{figure*}

Figures \ref{fig: case1} and \ref{fig: case2} show a graphical comparison of the uniform reference plan with locally optimal nonuniformly fractionated solutions for Cases 1 and 2; the graphical results for the remaining three cases can be found in the supplement. Table \ref{tbl:results} summarizes the computed bounds and benefits observed in all five cases. In each case, the nonuniform plan is constrained to maintain the same target coverage as the uniform reference plan, yet the mean liver BED in the nonuniform plans is substantially lower than in the uniform plans (see Table \ref{tbl:results}). Case 1 displays the smallest reduction in the mean liver BED with an approximately 13\% improvement, and Case 2 has the largest reduction of approximately 34\% in the mean liver BED. Note that these improvements are achieved without sacrificing any other clinical objective, as by definition, the spatiotemporal plans computed by solving \eqref{eq:IMRT-BED_constr} are at least as good as the uniform reference plan with respect to every objective. In particular, the spatiotemporal plans are as conformal as the corresponding uniform reference plans.

\begin{table*}
	\centering
	\small
\begin{tabular}{p{1ex} p{25ex} p{11ex} p{14ex} p{8ex} p{10.4ex} p{6ex} p{6ex}}
\toprule
& \multirow{2}{*}{Case description} & \multicolumn{4}{c}{Mean liver BED [Gy]} & Gap & Sparing \\ & & conventional  & spatiotemporal & reduction & lower bnd. & closed & factor\\
\midrule
	1 & Large central lesion & 84.54 & 75.87 & 12.75\% & 73.38 & 77.69\% & 0.6663 \\
	2 & Small lesion & 26.14 & 19.47 & 34.26\% & 18.58 & 88.23\% & 0.3830\\
	3 & Two small lesions &    59.54 & 50.24 & 18.51\% & 48.03 & 80.80\% & 0.5879\\
	4 & Lesion abutting ch.~wall & 47.51 & 38.65 & 22.92\% & 37.65 & 89.86\% & 0.5289 \\
	5 & Lesion abutting GI tract & 88.67 & 77.38 & 14.59\% & 77.02  & 96.91\% & 0.7028\\
\bottomrule
\end{tabular}
\caption{Summary of mean liver BED reductions from spatiotemporal fractionation, lower bounds for mean liver BED, and sparing factors for each of the five cases. The ``gap closed'' values provide a measure of how close the local optimal solutions are to achieving the lower bound on the mean liver BED; see Eq.~\eqref{eq:gap} for the definition. The sparing factor is a value that determines the dependence of the optimal fractionation schedule of a fixed dose distribution on the patient geometry (see Section \ref{sec:optimalNumFracs}). The remarkable benefit seen in Case 2 agrees with the fact that the sparing factor is substantially lower in this case than in the other cases.}
\label{tbl:results}
\end{table*}

The effectiveness of a spatiotemporal treatment can be quantified by its equivalent dose DEQ5, defined via
\begin{linenomath}
\begin{equation*}
\text{BED} = \sum_{t=1}^5 \left(d_t + \frac{d_t^2}{\ab} \right)
= \text{DEQ5} + \frac{(\text{DEQ5})^2/5}{\ab},
\end{equation*}
\end{linenomath}
which yields
\begin{linenomath}
\begin{equation*}
\text{DEQ5} = 5 \left( -\frac{\ab}{2} + \sqrt{\left( \frac{\ab}{2} \right)^2 + \frac{\ab}{5} \sum_{t=1}^5 \left(d_t + \frac{d_t^2}{\ab} \right)}\, \right).
\end{equation*}
\end{linenomath}
The DEQ5 is the cumulative dose distribution of a uniformly fractionated 5-fraction treatment that achieves the same BED as the given spatiotemporal treatment. (Note the difference between DEQ5 and the also commonly used \emph{equieffective dose}, EQX, which is the total physical dose that needs to be delivered in a uniform treatment with a dose per fraction of X Gy to achieve the same BED.) Using DEQ5 has the advantage that the spatiotemporal plan can be directly compared to the physical dose distribution of the uniform reference plan. The comparison is shown in panels (c)--(e) of Figures \ref{fig: case1} and \ref{fig: case2} for Cases 1 and 2, and in the Supplement for the remaining cases. By computing the difference between the two plans, we observe that the nonuniformly fractionated plans maintain the same tumor BED as the uniformly fractionated plans while reducing the mean dose and BED in healthy liver tissue. The same can also be demonstrated by comparing the DVH curves of the physical dose and DEQ5 of the two plans. For Case 1, these curves are included in the Supplement (Figure 7). The DVH curves show that there is substantial dose reduction in the liver (both the DEQ5 and physical dose curves shift to the left), and that this is a consequence of lower physical dose (but nearly identical DEQ5) delivered to the tumor.

\subsubsection{Bounds on the maximum achievable liver BED reduction.}
Because the nonuniform fractionation problem is nonconvex, the nonuniformly fractionated solutions presented in this paper are only certified to be locally optimal. To gauge the quality of these solutions, for each case we computed a lower bound on the minimum liver BED, as described in Section~\ref{sec:nonuniform-constrained}. These lower bounds, also shown in Table \ref{tbl:results}, are close to the mean liver BED values achieved by the locally optimal nonuniformly fractionated solutions. For example, in Case 1, spatiotemporal fractionation reduced the mean liver BED to 75.9 Gy, which is a reduction of 8.7 Gy from the uniform reference plan. The lower bound from the SDP relaxation is 73.4 Gy, which means that no treatment plan can achieve a reduction of more than 11.2 Gy from the uniform reference plan. It is important to note that the value of 73.4 Gy is only a lower bound, and we are not able to guarantee the existence of a nonuniformly fractionated solution that achieves this value. The mean liver BED value for the true globally optimal plan may be anywhere between 75.9 and 73.4 Gy.

We compared the mean liver BED reduction in the locally optimal nonuniformly fractionated plan with our bound on the maximum possible reduction by computing the fraction of the gap between the mean liver BED in the uniform reference plan and the lower bound from the SDP relaxation that the nonuniformly fractionated plan closes. Formally, if $m_\text{ref}$ and $m_\text{sp}$ are the mean liver BEDs achieved by the uniform reference plan and the locally optimal spatiotemporal plan, respectively, and $m_\text{SDP}$ is the lower bound obtained from the SDP relaxation, then the fraction of the \emph{gap closed} by the spatiotemporal plan is the ratio
\begin{linenomath}
\begin{equation}\label{eq:gap}
\frac{m_\text{ref}-m_\text{sp}}{m_\text{ref}-m_\text{SDP}}.
\end{equation}
\end{linenomath}
In particular, a ratio of $100\%$ is the best possible value, which is attained if the spatiotemporal plan matches the SDP lower bound, and $0\%$ is the worst possible value, attained when the uniform plan cannot be improved upon using spatiotemporal fractionation. Note that this ratio being close to $100\%$ does not only speak directly for the quality of the spatiotemporal plan; it also demonstrates the sharpness of the lower bound. For example, in Case 1, the nonuniform plan's reduction of 8.7 Gy is 78\% of the upper bound on the maximum possible BED reduction, which is 11.2 Gy. The nonuniformly fractionated plans closed 78-97\% of the gap between the mean liver BED of the uniform reference plan and our lower bound on the lowest achievable mean liver BED.

\subsubsection{The optimal number of fractions.}\label{sec:optimalNumFracs}
In order to fairly evaluate the benefit that comes from spatiotemporal fractionation, we need to consider the gain to be had from changing the fractionation schedule without altering the physical dose distribution. In particular, some cases might already benefit from hypofractionation alone, in which case the improvement we see in the spatiotemporal plans could be a combination of the benefit of allowing for greater hypofractionation in the tumor and the benefit of allowing for different dose distributions in different fractions. We computed the optimal uniformly fractionated treatment plans with up to five fractions by solving the uniform model \eqref{eq:IMRT-BED_uni} for  $N=1,\dots,5$. In all of the cases, the five-fraction uniform plans had the lowest mean liver BED among all computed uniform plans. Hence, Table \ref{tbl:results} compares the 5-fraction spatiotemporal plans against the best uniform plans with up to five fractions.

This agrees with previous work on the dependence of optimal fractionation schedules on the patient geometry and dose distribution. When the goal is to minimize the mean dose to a single dose-limiting parallel organ such as the liver, the optimal fractionation schedule is a function of the \emph{effective sparing factor} $\bar{\delta}$ which was introduced independently by \citeasnoun{UnkelbachCraftSalariRamakrishnanBortfeld2013} and \citeasnoun{KellerHopeMeierDavison2013}. It was shown that if $\bar{\delta} > \frac{\left(\ab\right)_N}{\left(\ab\right)_T} = 0.4$, then increasing the number of fractions is optimal, and if $\bar{\delta} < 0.4$, then lowering the number of fractions is optimal. In Case 2, the effective sparing factor is approximately $0.4$, while in the remaining four cases, the sparing factor is well above $0.4$. This provides an explanation for the observation that the benefit of spatiotemporal fractionation was largest for Case 2; the treatment quality of a uniformly fractionated 5-fraction treatment and a single-fraction treatment is approximately equal. Therefore, achieving a benefit through spatiotemporal fractionation relies to a lesser extent on achieving near-uniform fractionation in the normal liver.

\section{Discussion}\label{sec:discussion}

\paragraph{Results.} The results indicate that spatiotemporal treatments can achieve substantial reductions in both BED and physical dose to the liver and other normal tissue. The approximately 15-20\% mean liver BED reduction is consistent with the benefit of spatiotemporal fractionation observed in previous work for brain lesions \cite{UnkelbachBussiereChapmanLoefflerShih2016}.  Our work further shows that local optimization techniques provide high-quality plans that are close to realizing the maximum potential normal tissue dose reduction.

The results in this paper are limited to the optimization of treatment plans for two-dimensional slices of clinical liver cases in order to demonstrate the concepts introduced in the paper, but in a practical setting the computations would have to be carried out for the entire patient. There is no additional difficulty in computing locally optimal nonuniform solutions for three-dimensional cases with the same algorithms that we used in our study  \cite{UnkelbachBussiereChapmanLoefflerShih2016}. Therefore, the main challenge in extending this study to three-dimensional cases is the computational complexity of solving the SDP relaxations in Section \ref{sec:SDP-bound}. In the nonuniform fractionation optimization problem there is at least one variable for each beamlet in the fluence maps and at least one constraint for every voxel associated with a quadratic penalty function. In the SDP relaxation, with the introduction of the matrix variables $X$, the number of variables is roughly squared. SDP problems of this size cannot be solved for three-dimensional cases in a reasonable amount of time with the available off-the-shelf software. Even with the two-dimensional case, the solver took up to 7 hours to solve the SDP relaxation. It is an interesting direction for future research to devise customized algorithms for the solution of these large-scale SDP problems that arise for the three-dimensional cases. For now, the results obtained for the two-dimensional slices suggest that the locally optimal solutions computed for the nonuniform fractionation problem are close to global optimality.

\paragraph{Static-beam IMRT versus VMAT.} As mentioned in Section \ref{sec:results}, the treatment plans in this work have a higher number of beams than a conventional IMRT plan to approximate a VMAT plan, where we expect nonuniform fractionation to display the most benefit. The nonuniformly fractionated plans that are most effective in lowering BED and total physical dose are those in which the tumor is hypofractionated while the surrounding tissue is fractionated. In other words, these plans deliver a high single-fraction dose to parts of the tumor during each fraction and a consistent lower dose to the liver and other healthy tissue throughout all of the fractions. VMAT treatments are particularly suitable for delivering such a nonuniformly fractionated plan, thanks to their characteristic low-dose bath delivered to the healthy tissue.

\paragraph{Uncertainty.}
Further research is needed to investigate the impact of various sources of uncertainty on spatiotemporally fractionated treatments. We anticipate that the effect of small changes in the radiobiological parameters is negligible, as they only slightly affect what fractionated treatments are isoeffective. Interfractional patient motion and soft tissue deformation is a greater concern for spatiotemporal treatments in which each fraction delivers precariously aligned dose distributions with sharp dose gradients in the interior of the target structure. Future research will consider including additional safeguards such as bounds on the dose gradients as well as stochastic and robust optimization methods to explicitly incorporate uncertainty into the planning. We also note that spatiotemporal fractionation can be applied to treatments with fewer distinct fractions that are delivered a number of times each; for example, a 15-fraction scheme can be designed (entirely analogously to our plans) that consist of only 3 distinct fractions that are delivered 5 times each. The repeated delivery of identical fractions may mitigate somewhat the effects of random interfraction motion. Finally, other technologies, currently under development, may also make the safe delivery of spatiotemporal liver SBRT treatments feasible in the future. These include MLC and couch tracking, and image-guided radiotherapy using an MR-linac.

\ack
This material is based upon work supported by the National Science Foundation under Grant No.~DMS-1719828. Additionally, this material was based upon work partially supported by the National Science Foundation under Grant DMS-1127914 to the Statistical and Applied Mathematical Sciences Institute. Any opinions, findings, and conclusions or recommendations expressed in this material are those of the authors and do not necessarily reflect the views of the National Science Foundation.

\section*{Disclosures}

\noindent The authors have no conflicts of interest to disclose.

\bigskip

\section*{References}
\bibliography{STFracWithOptimalityBounds_bib_arXiv}

%
%
%

\clearpage

\section*{Supplementary Figures}

\begin{center}
\setcounter{figure}{3}
\begin{figure*}[h]
\makebox[\textwidth][c]{
\begin{tabular}{c c c c}
\multicolumn{4}{c}{(a)} \\
\multicolumn{4}{c}{\includegraphics[width=\textwidth]{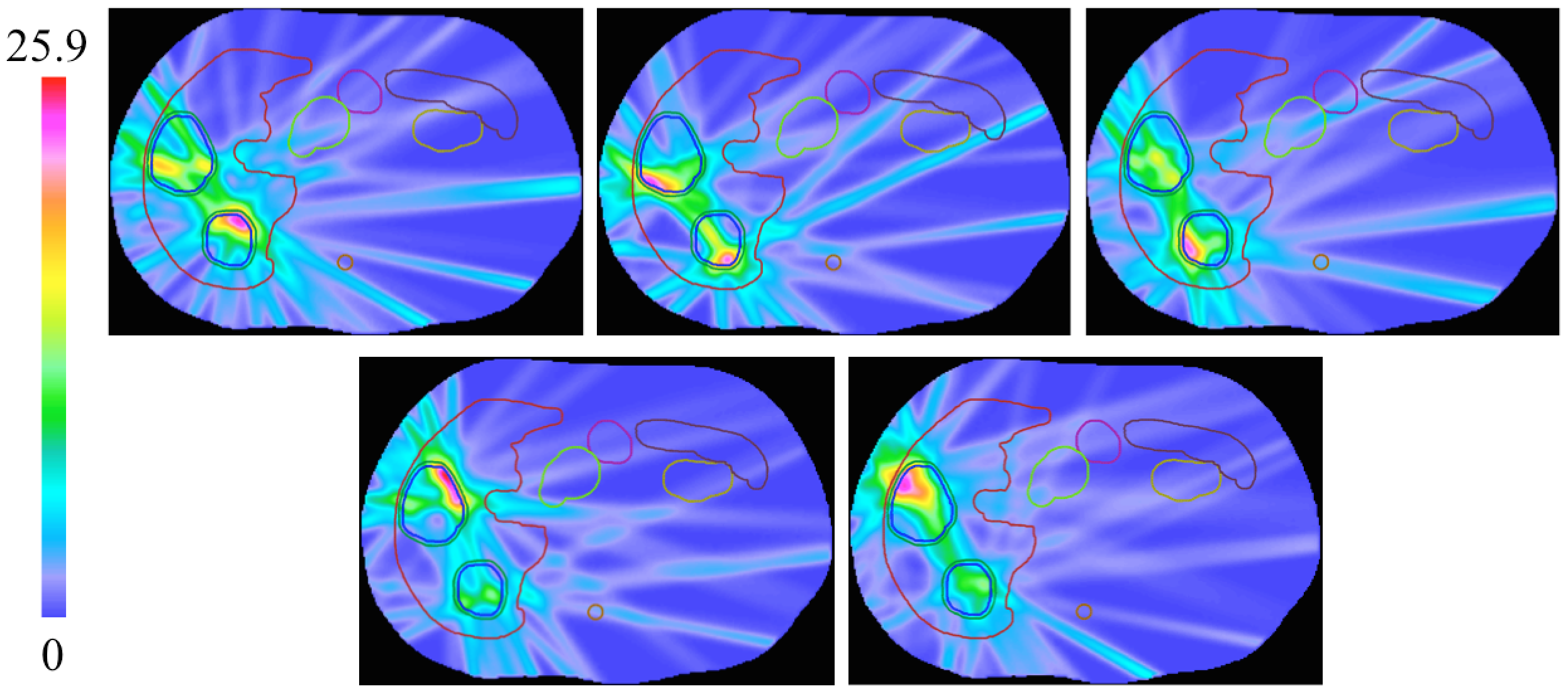}} \\
(b) & (c) & (d) & (e) \\
\includegraphics[height=7em]{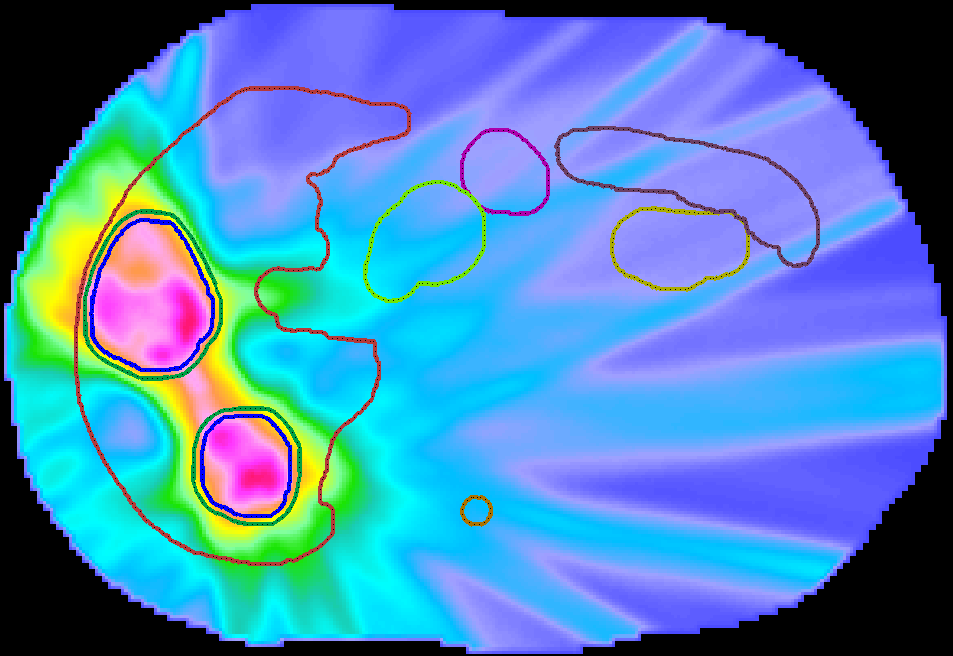} &
\includegraphics[height=7em]{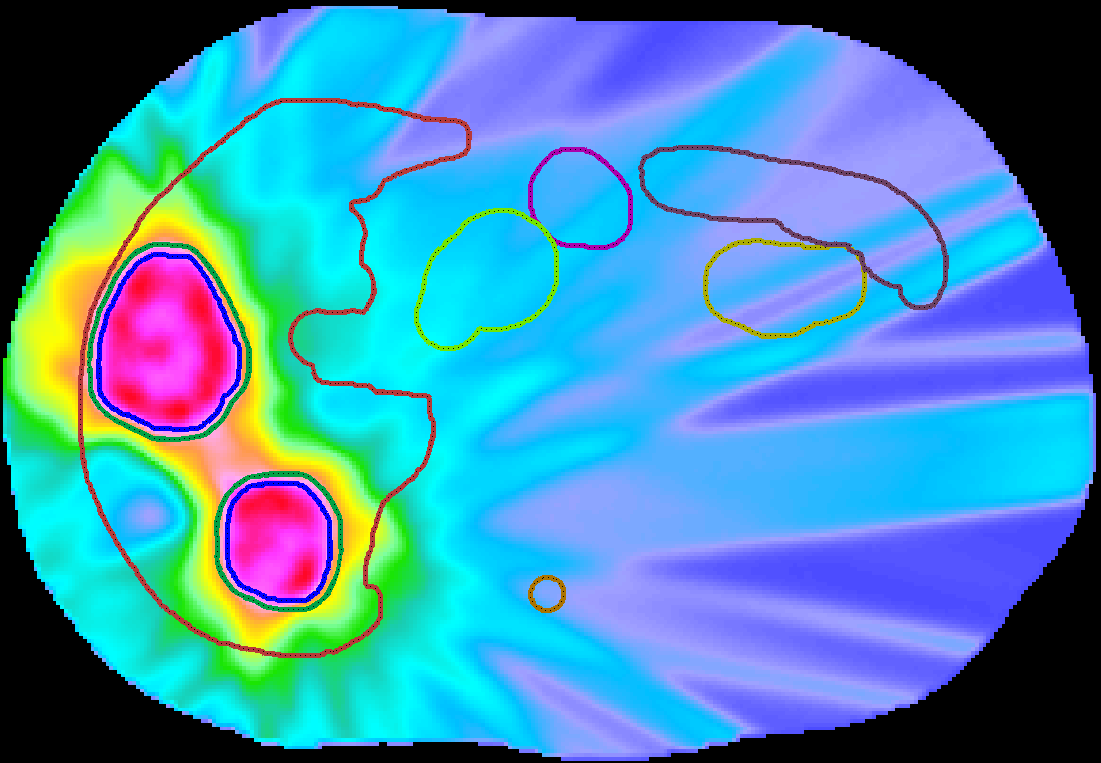} &
\includegraphics[height=7em]{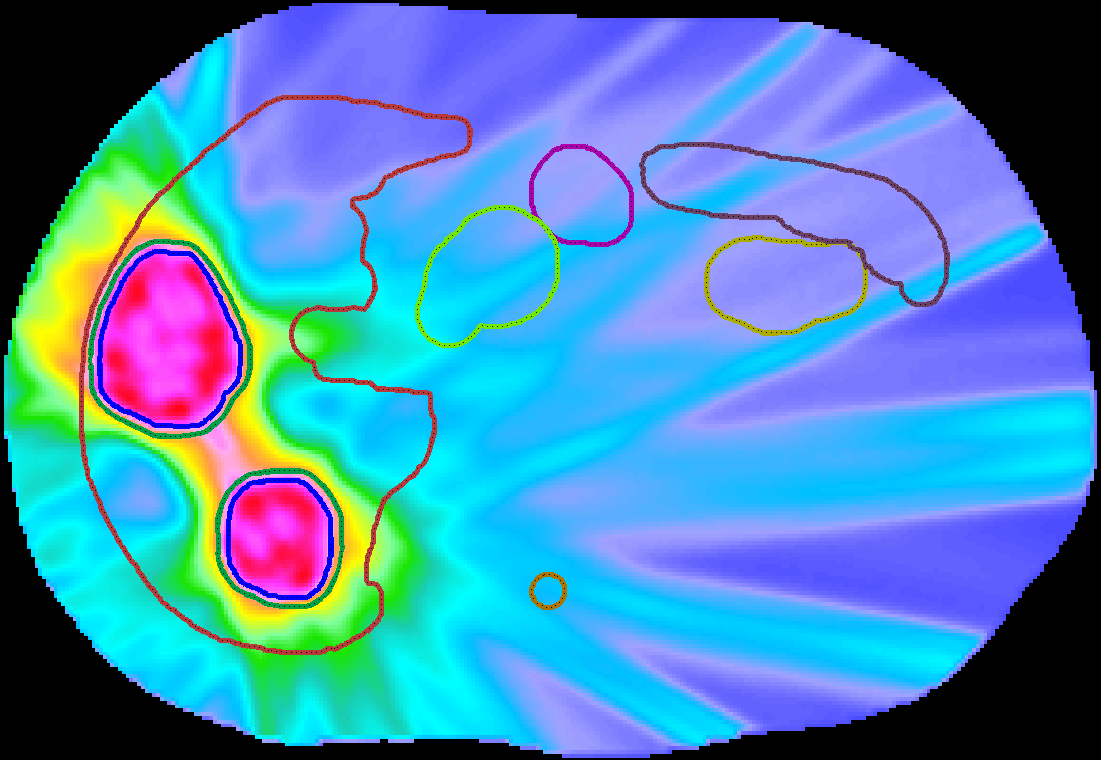} &
\includegraphics[height=7em]{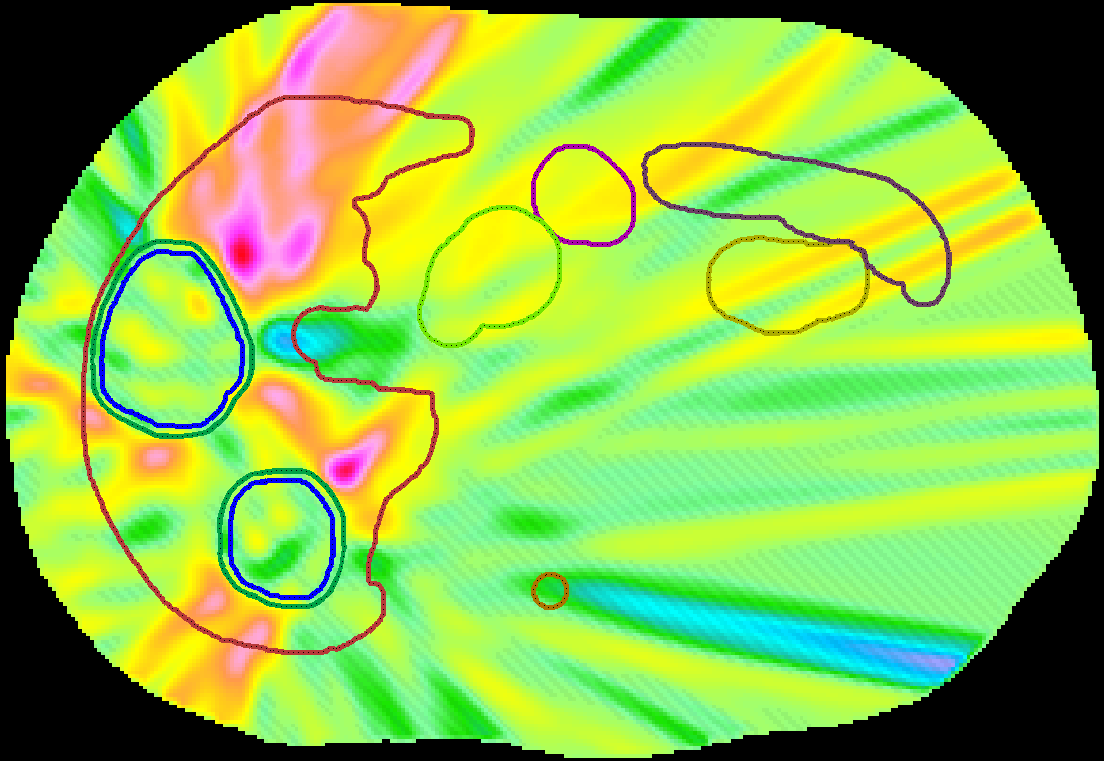} \\
\multicolumn{3}{c}{\includegraphics[height=.5cm]{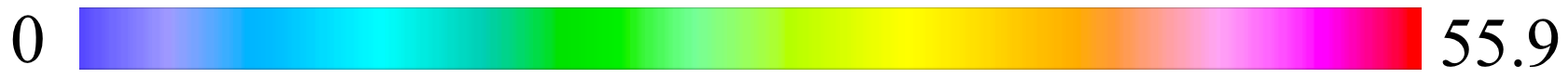}} &
\includegraphics[height=.45cm]{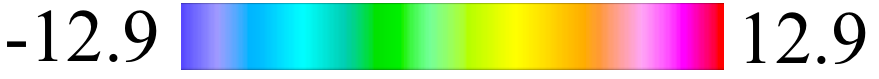} \\
\end{tabular}
}
\caption{Nonuniformly-fractionated dose distributions for Case 3, which contains two lesions within the liver. (a) Physical dose distributions in each of the five fractions show that the nonuniformly fractionated treatment hypofractionates different parts of the tumor. (b) Total physical dose delivered throughout the nonuniformly fractionated treatment. (c) Physical dose distribution of the uniformly fractionated reference plan. (d) DEQ5 of the nonuniformly fractionated plan, which is the uniform plan that is isoeffective in delivering the same BED as the nonuniformly fractionated plan. (e) The difference between the physical dose in the uniform plan and the DEQ5 for the nonuniform plan, or (c) minus (d). This shows that the benefit of nonuniform fractionation is mostly in reduced dose in the healthy liver and in the entrance region of the those beams that expose the liver the most. All numerical quantities shown are in [Gy].}
\label{fig: case1}
\end{figure*}

\begin{figure*}[h]
\makebox[\textwidth][c]{
\begin{tabular}{c c c c}
\multicolumn{4}{c}{(a)} \\
\multicolumn{4}{c}{\includegraphics[width=\textwidth]{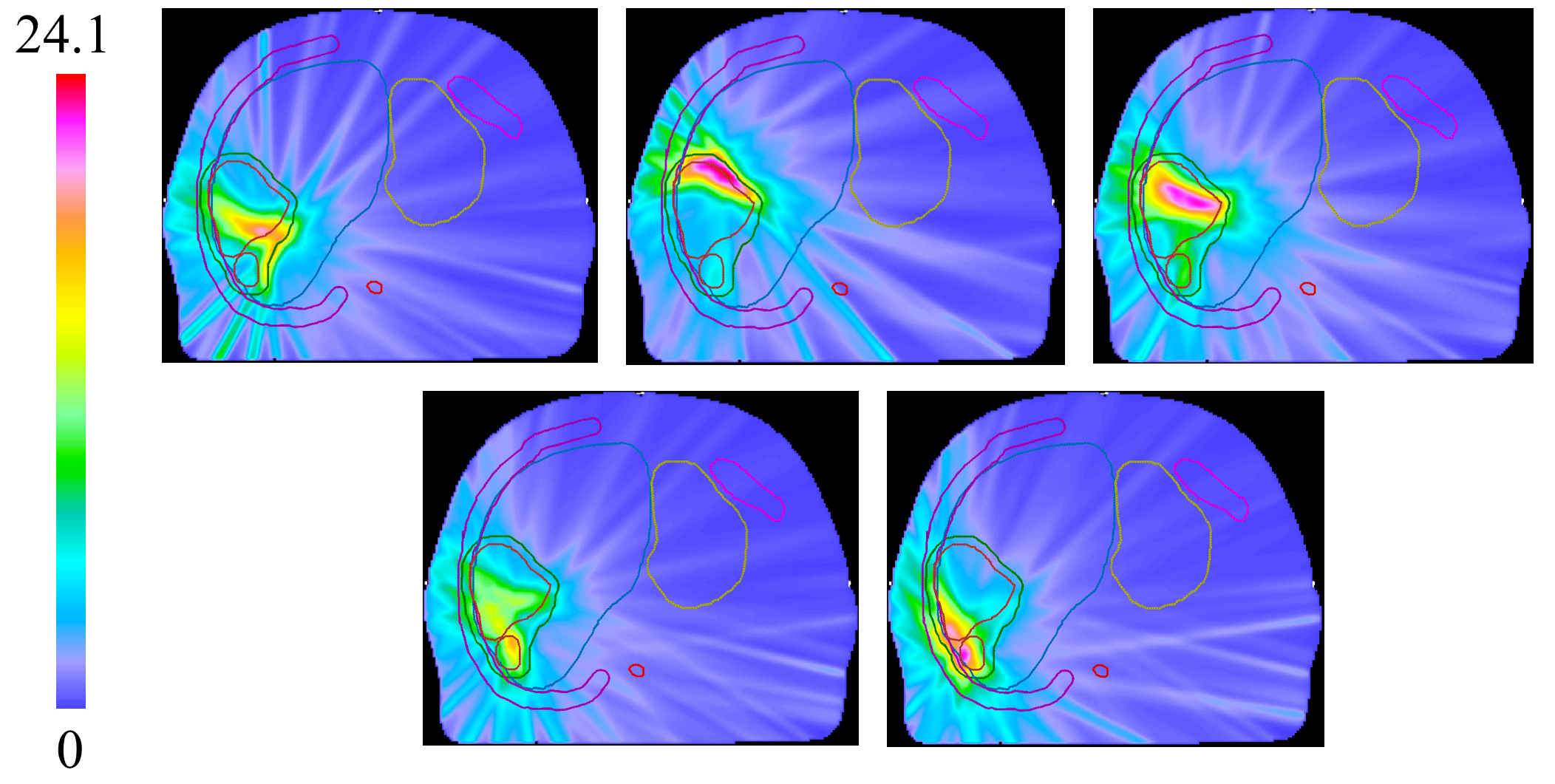}} \\
(b) & (c) & (d) & (e) \\
\includegraphics[height=8em]{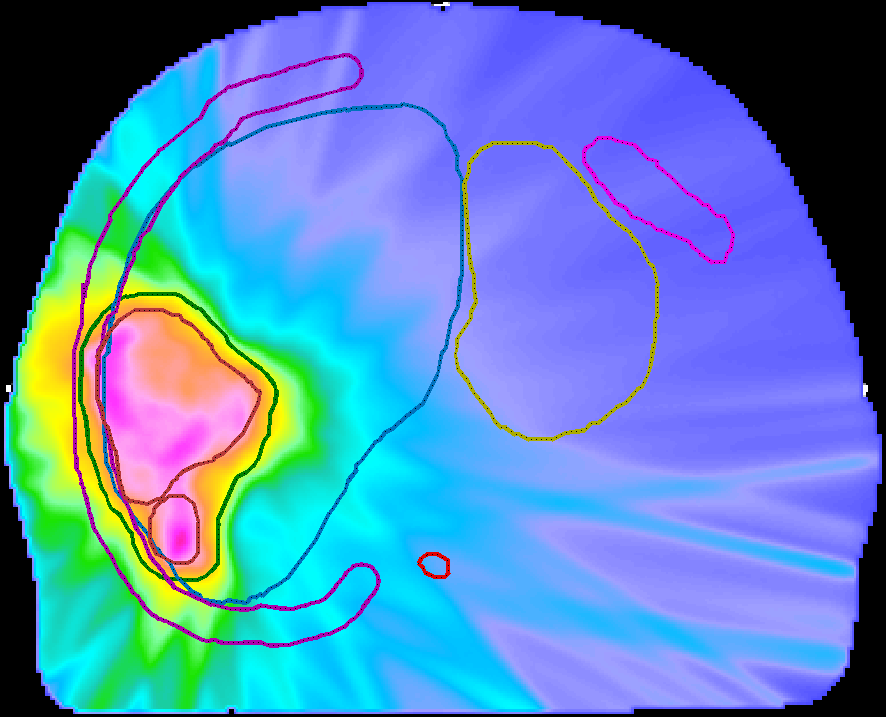} &
\includegraphics[height=8em]{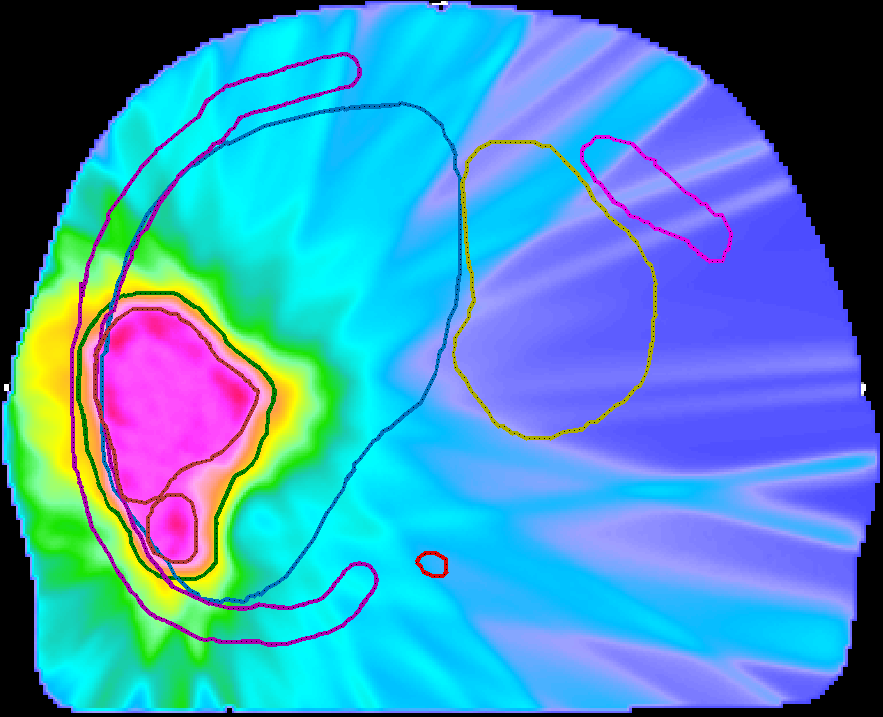} &
\includegraphics[height=8em]{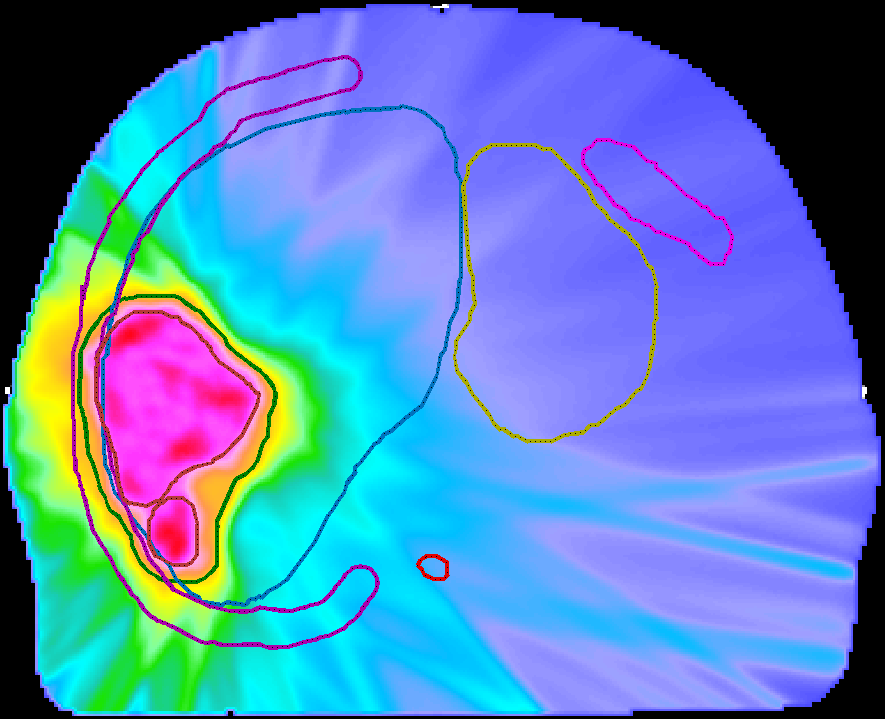} &
\includegraphics[height=8em]{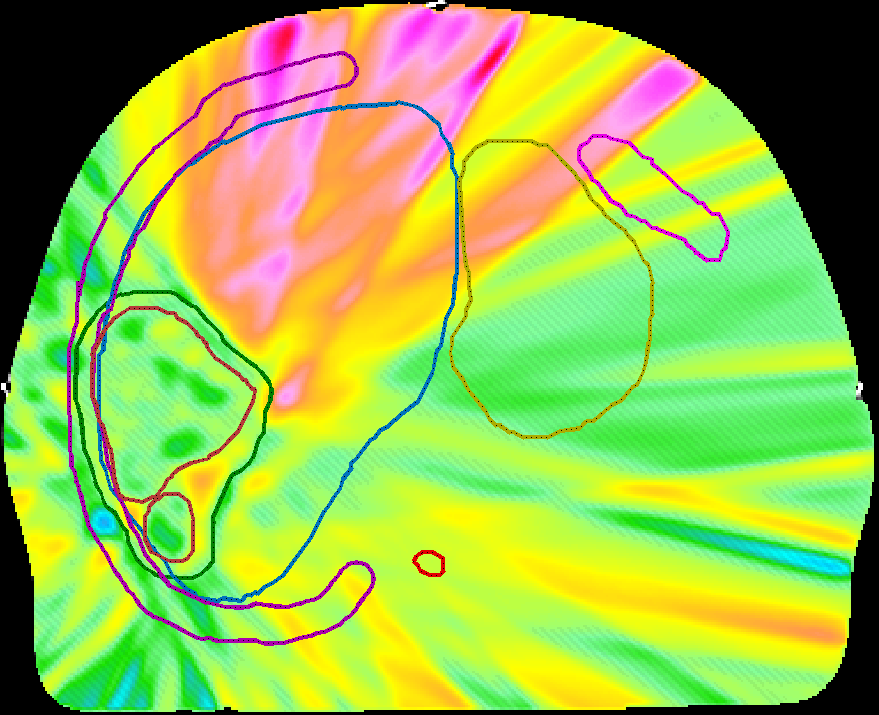} \\
\multicolumn{3}{c}{\includegraphics[height=.6cm]{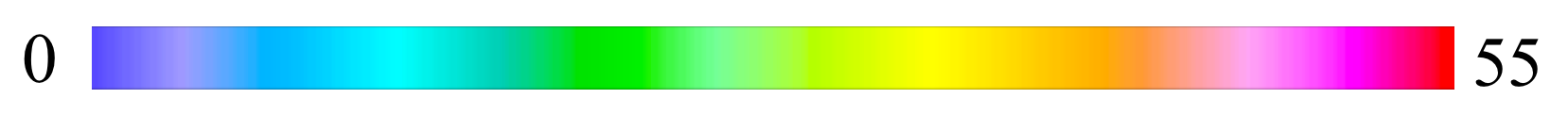}} &
\includegraphics[height=.5cm]{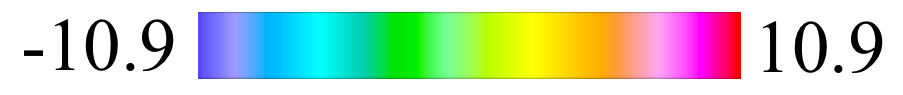} \\
\end{tabular}
}
\caption{Dose distributions for Case 4, a lesion is abutting the chest wall. (a) Physical dose distributions for each fraction of a nonuniformly fractionated treatment. (b) Total physical dose delivered by the five nonuniform fractions. (c) Physical dose distribution of the uniform reference plan. (d) DEQ5 of the nonuniformly fractionated plan. (e) Plot of (c) minus (d). All numerical quantities shown are in [Gy].}
\label{fig: case4}
\end{figure*}

\begin{figure*}[h]
\makebox[\textwidth][c]{
\begin{tabular}{c c c c}
\multicolumn{4}{c}{(a)} \\
\multicolumn{4}{c}{\includegraphics[width=\textwidth]{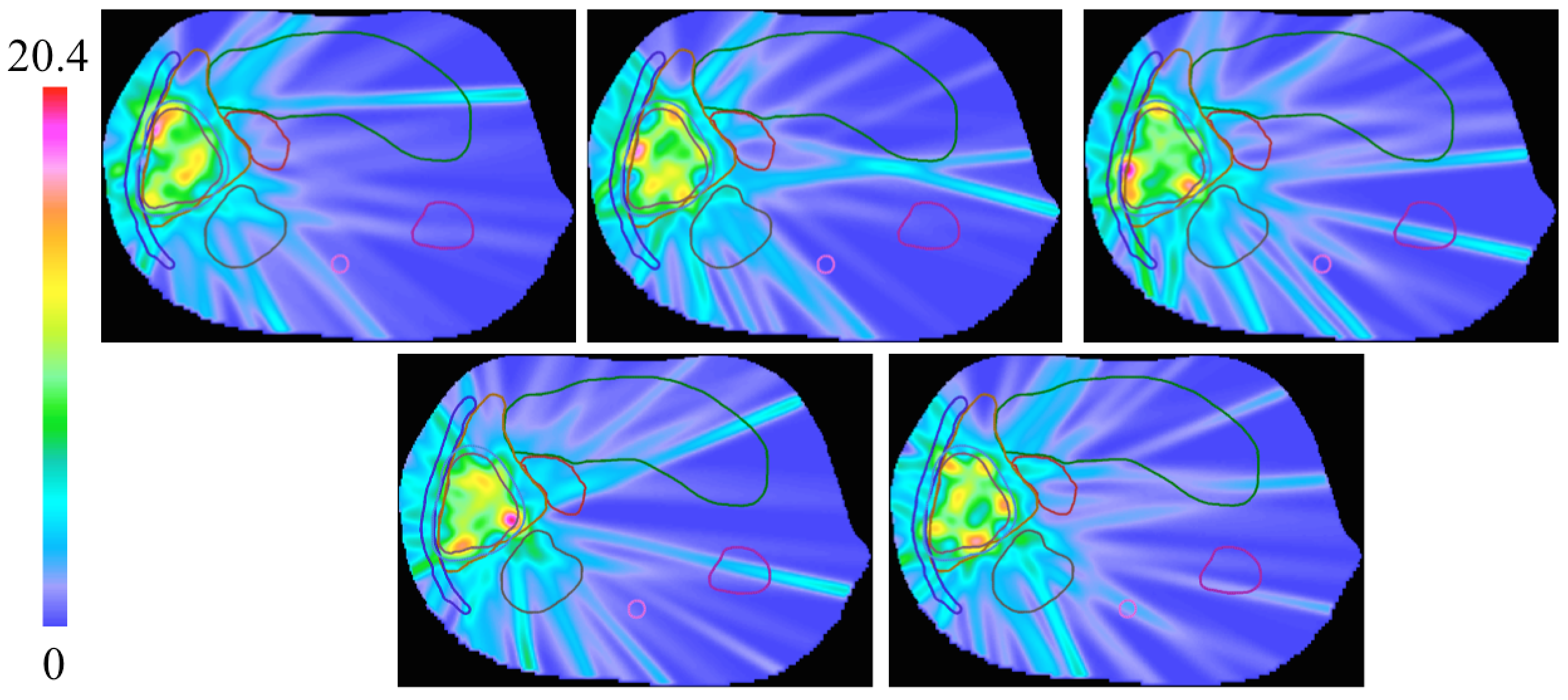}} \\
(b) & (c) & (d) & (e) \\
\includegraphics[height=7em]{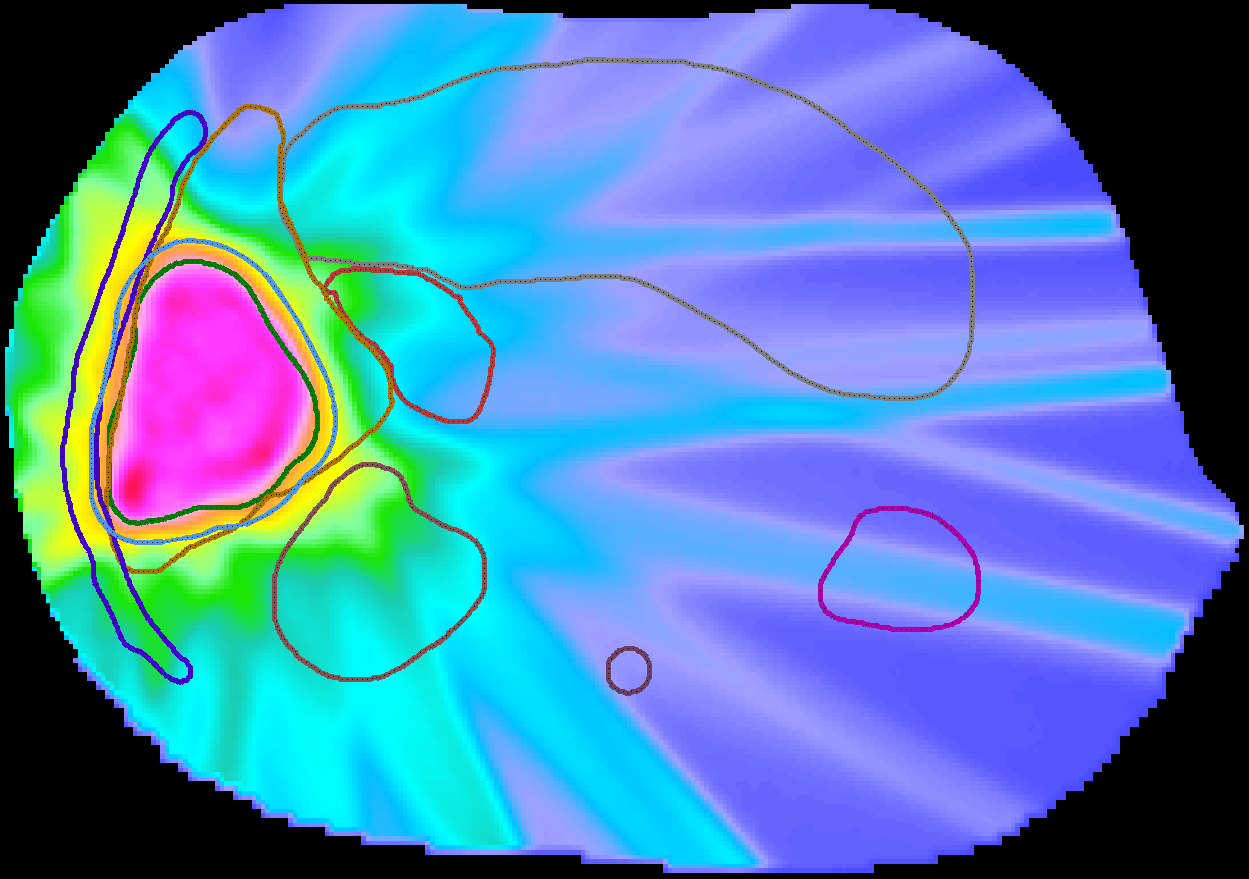} &
\includegraphics[height=7em]{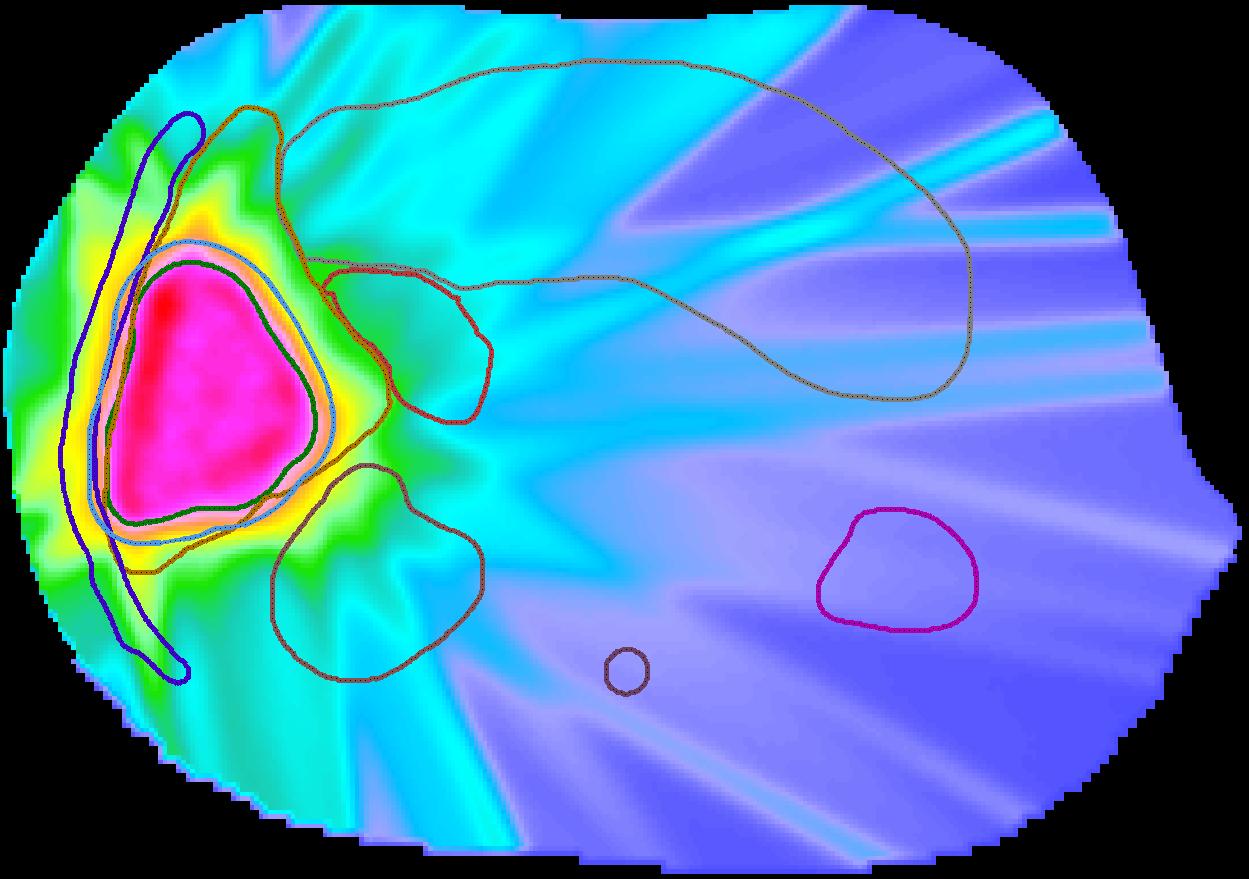} &
\includegraphics[height=7em]{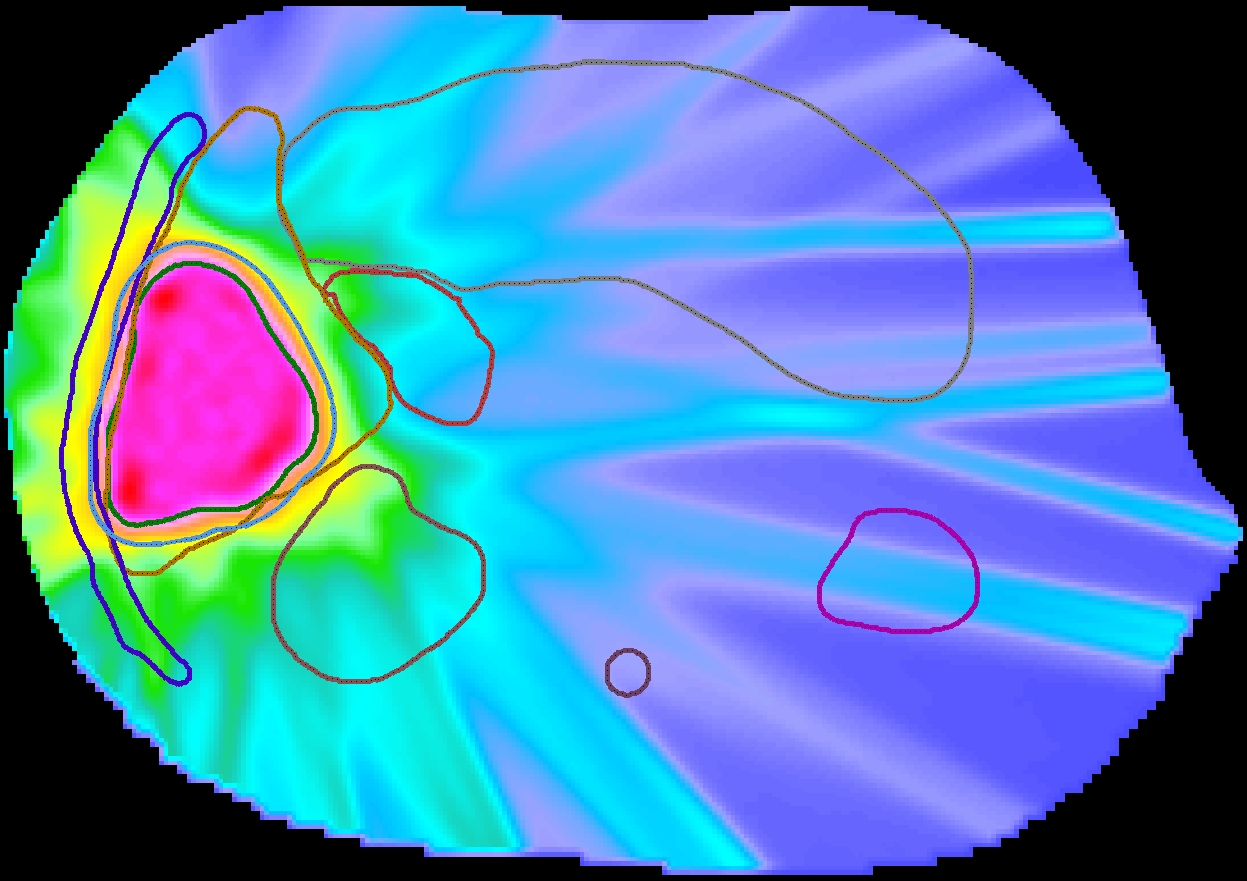} &
\includegraphics[height=7em]{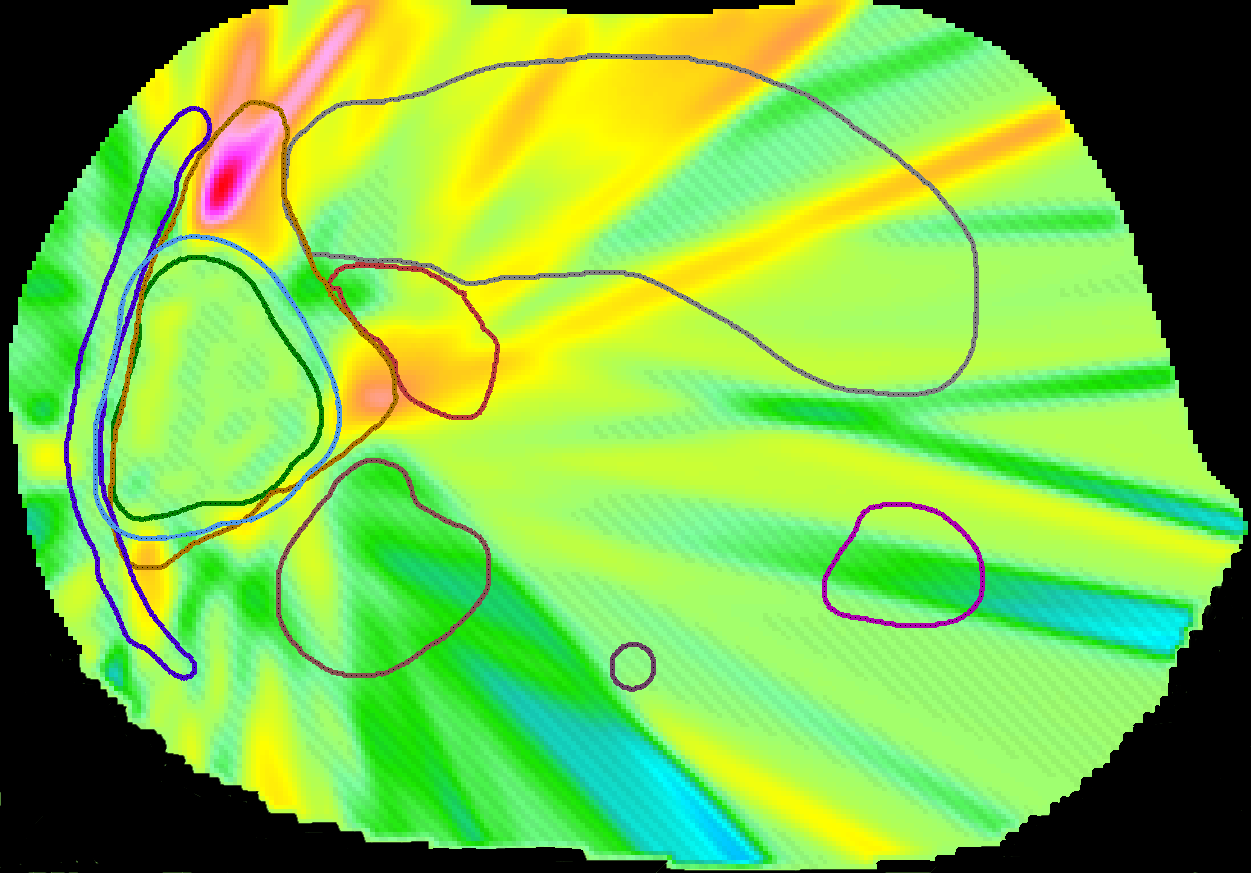} \\
\multicolumn{3}{c}{\includegraphics[height=.5cm]{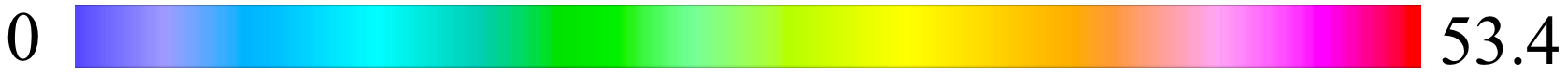}} &
\includegraphics[height=.45cm]{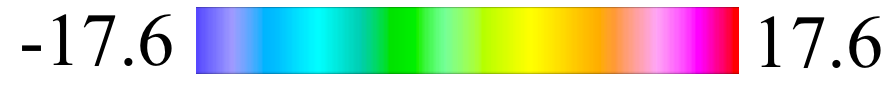} \\
\end{tabular}
}
\caption{Dose distributions for Case 5, a lesion abutting the GI tract. (a) Physical dose distributions for each fraction of a nonuniformly fractionated treatment. (b) Total physical dose delivered by the five nonuniform fractions. (c) Physical dose distribution of the uniformly fractionated reference plan. (d) DEQ5 of the nonuniformly fractionated plan. (e) Plot of (c) minus (d). All numerical quantities shown are in [Gy].}
\label{fig: case5}
\end{figure*}

\begin{figure*}[h]
\centering
\begin{tabular}{c}
\includegraphics[width=.9\textwidth]{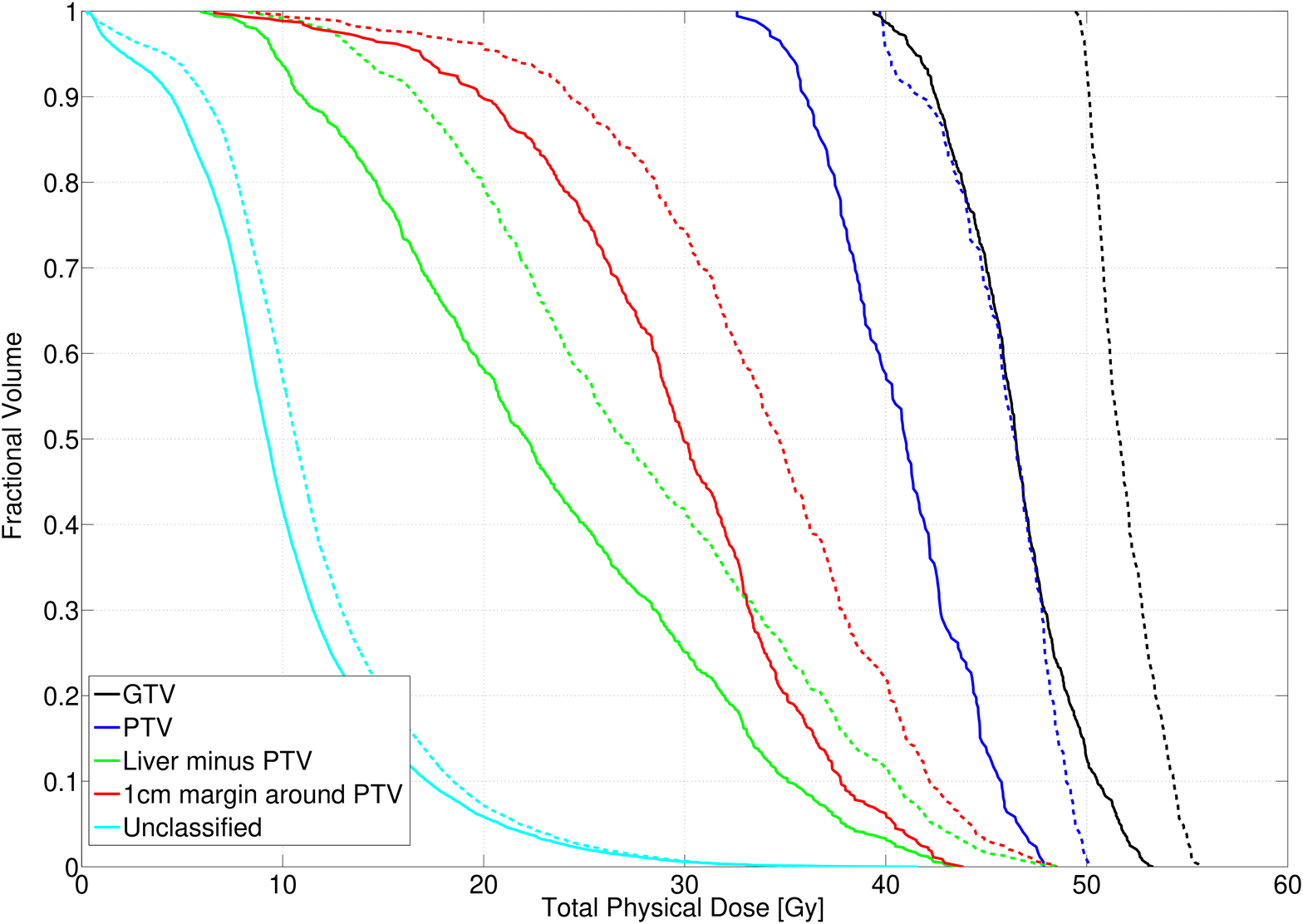} \\
(a) \\
\includegraphics[width=.9\textwidth]{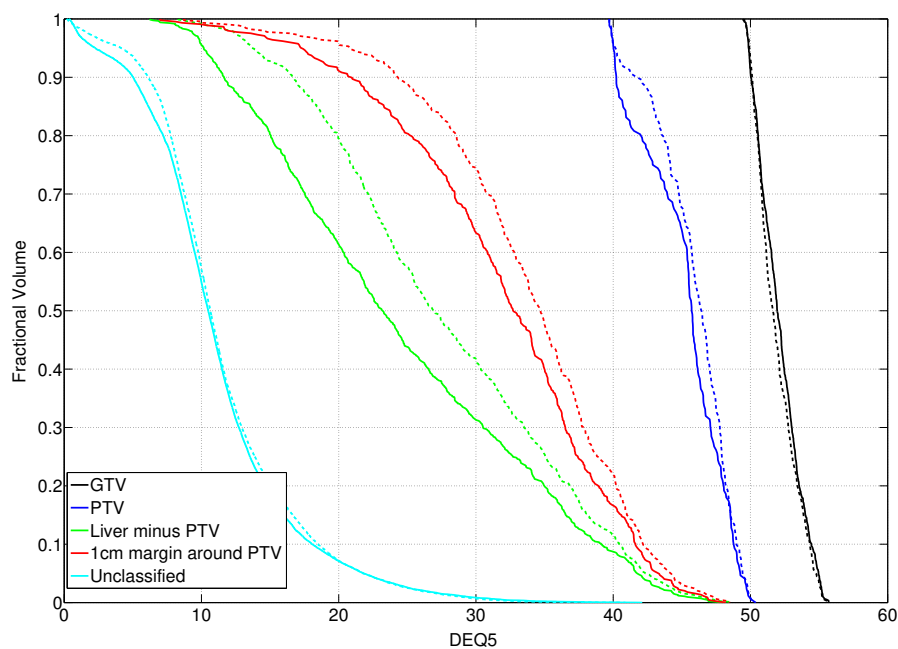} \\
(b)
\end{tabular}
\caption{Dose-volume histogram for (a) total physical dose and (b) DEQ5 for various structures in Case 1. The dotted lines are the values from the uniformly fractionated reference plan and the solid lines are from the spatiotemporal plan. These curves indicate that the spatiotemporal plan achieves an overall reduction in physical dose, as observed by the lines shifted to the left in (a). Additionally, in (b) we note that the spatiotemporal plan maintains DEQ5 in the tumor while reducing dose to the healthy liver tissue.}
\label{fig:DVHs}
\end{figure*}

\end{center}


\end{document}